*Inaugural Keynote Lecture · 31st Col·loquis de Vic, Forgetting (L'oblit) · Vic, 8–9 October 2026*

# Prefigured Forgetting

## Memory, the Right to Be Forgotten and Artificial Intelligence

**POMPEU CASANOVAS**

*Institut de Recerca en Intel·ligència Artificial (IIIA-CSIC)*
*La Trobe University*
*Institut de Dret i Tecnologia (IDT-UAB)*
*pompeu.casanovas@iiia.csic.es · https://orcid.org/0000-0002-0980-2371*

**Abstract.** This inaugural keynote lecture surveys the state of the art of the right to be forgotten and examines how it is being transformed by artificial intelligence. Drawing on cognitive science, philosophy and the social sciences, it treats forgetting as a constitutive function of memory, and recalls that remembering the past and imagining the future share the same mechanism. It then reviews the origins, the European and national case law, and the paradoxes of the right, and offers a periodisation of the specialised literature (2009–2026). The core claim is that digital memory has moved through three regimes—archive, search engine and generative model—and that the right was designed for the first two. In the third, personal information is no longer stored as a locatable record but inferred and regenerated; AI prefigures the frameworks of individual and collective remembering. The right to be forgotten thus ceases to be merely an individual *ex post* remedy and becomes a property to be designed and governed within legal and ethical ecosystems along the AI value chain.

## 1 Introduction: from the imagined future to prefigured forgetting

Last year, in these Col·loquis, we asked how we represent the future. I argued then that images of what is to come—Junceda's cartoons in *En Patufet* (1926), Côté's postcards, the Echte Wagner trade cards—are bearers of values, have a collective subject and are anchored in the everyday world of those who produce them. The limit of our capacity to prefigure, I said, is that "we are guided by what we can see", and that our perceptions are constrained by symbolic phantasms close at hand.[1] This year the theme is forgetting. It may look like a complete about-turn—from tomorrow to yesterday—but I would like to show that it is the same problem seen from the other side.

This is so, first, for a cognitive reason. Research on episodic memory has established that remembering the past and imagining the future rely to a large extent on the same system: memory is not an archive but a constructive process that recombines fragments of experience, and it is precisely this flexibility that allows us to simulate scenarios that have not yet happened.[2] Karl Bühler had seen this from the standpoint of language: *Deixis am Phantasma*, the third mode of deixis alongside *demonstratio ad oculos* and anaphora, points equally to what is remembered and to what is fantasised, because in both cases the speaker moves into a

1 Casanovas (2026), Sect. 5.

2 Schacter and Addis (2007).

field that is not before them.[3] If this is so, whoever intervenes in the raw material of memories also intervenes in the repertoire of the imaginable future. What a society can expect depends on what it can remember, and on what it is allowed to forget.

Second, it is so for a biographical reason that I hope I may be allowed to mention. In 2007 the Col·loquis were devoted to memory, and I presented a paper on written memory and graphic images.[4] Nineteen years later, the circle closes with its reverse. In between, something has happened that we could not then foresee: the default condition has been inverted. For millennia, forgetting was what happened if nothing was done, and remembering required effort, technique and institutions. Today preservation is free and automatic, and it is forgetting that demands a decision, a cost and—this is our subject—a right.[5]

The right to be forgotten is the European legal response to this inversion. It was born in an environment of documents and archives, was consolidated in relation to search engines, and now confronts a technology that fits neither mould. Generative artificial intelligence models, I said last year, introduce probability "as a factual element in our way of feeling, thinking, arguing and deciding".[6] We must now add: also in our way of remembering and of being remembered. A language model does not preserve a news item about me; it has learnt statistical regularities from it and, when someone asks, generates a plausible one. It does not evoke: it prefigures.

I use this verb deliberately. To prefigure is to configure in advance: to arrange the environment, the context and the frameworks within which something will be able to appear. My thesis is that AI now has the capacity to prefigure memory—individual and collective—in three senses: it *selects* what becomes visible of the past, it *regenerates* what is no longer there (or never was), and it *feeds back* its products into human memories and into the data of future models. I further contend that this affects the possibility of effectively exercising the right to be forgotten, because the right presupposes an object—a datum, a link—and a controller, both of which dissolve in the new regime.

I shall proceed in five steps. I shall review what we know about forgetting from cognitive science, philosophy and the social sciences (Sect. 2); reconstruct the origins, the case law and the paradoxes of the right to be forgotten (Sect. 3); propose a periodisation of the specialised literature (Sect. 4); analyse what happens to the right when it comes into contact with AI (Sect. 5); and conclude by asking what we can expect (Sect. 6). The common thread is the one I have developed in earlier work on legal governance and smart legal and ethical ecosystems: in a hybrid society, rights are guaranteed not only by norms and judgments, but by the design of the environments in which humans and intelligent systems interact.[7]

3 Bühler (1934), § 8. See Casanovas (2026), Sect. 2.

4 Casanovas (2007).

5 Blanchette and Johnson (2002); Mayer-Schönberger (2009), pp. 1–15; Esposito (2017), p. 5.

6 Casanovas (2026), Sect. 2.

7 Casanovas (2024); Casanovas and Noriega (2026).

## 2 The concept of forgetting in cognitive science, philosophy and the social sciences

### 2.1 Forgetting and memory, individual and collective

Experimental psychology began measuring forgetting before it measured memory. Ebbinghaus's curve (1885) described loss as a function of time; Bartlett (1932) showed that what remains is not a diminished copy but a reconstruction guided by cultural schemata.[8] For a century, however, forgetting was treated as a passive failure: decay or interference. The change of perspective is recent and clear-cut. Anderson and Hulbert have documented mechanisms of prefrontal control that actively suppress competing memories: we forget, in part, because the brain does so, and sometimes because we want to.[9] At the 2022 Col·loquis I gave examples of cognitive impenetrability as protection, that is, of forgetting as a survival mechanism.[10] John J. Drummond, for instance, recounts how as a child he fell onto a bees' nest and was stung more than seventy times: the fear and the pain endure, while the details of the episode have been erased from his memory.[11]

Neurobiology has shown that the function of memory is not to transmit information faithfully through time but to optimise decision-making in changing environments, and that transience serves this end in two ways: it increases flexibility, by reducing the weight of outdated information, and it prevents overfitting to particular episodes, thereby favouring generalisation.[12] The vocabulary is that of machine learning, and not by chance. We shall return to this.

Literature already knew this. Borges's Funes, incapable of forgetting, was also incapable of thinking, because "pensar es olvidar diferencias, es generalizar, abstraer" [to think is to forget differences, to generalise, to abstract]; and the mnemonist whom Aleksandr Luria followed for thirty years, Solomon V. Shereshevsky, lived besieged by a memory that could let go of nothing, sustained by images that he placed mentally in a fixed order, as in the ancient method of loci.[13] Schacter has systematised the "seven sins" of memory—transience, absent-mindedness, blocking, misattribution, suggestibility, bias and persistence—as by-products of adaptive traits: the price of a system that works.[14] Let us keep two of them in mind, misattribution and suggestibility, because they are precisely the ones that conversational AI is capable of exploiting (Sect. 5.3).

Philosophy has conceived of forgetting both as a threat and as a condition. In the *Phaedrus*, King Thamus warns that writing is a *phármakon* (remedy and poison at once) that will produce forgetfulness in the souls of those who rely on it; it is the first critique of the externalisation of memory, and we have repeated it with every technology. The experience of Australian Aboriginal peoples, who have transmitted memory and knowledge over millennia

8 Ebbinghaus (1885); Bartlett (1932).

9 Anderson and Hulbert (2021).

10 Casanovas (2023).

11 Drummond (2004), pp. 109–110.

12 Richards and Frankland (2017).

13 Borges (1944); Luria (1968).

14 Schacter (2001).

without writing, through symbols, stories, songs and pictograms, shows that a memory without writing can be extraordinarily durable—which, in a sense, bears Thamus out.[15] Nietzsche reversed the sign: without the capacity to forget there is neither action nor happiness, and an excess of history stifles life.[16] Ricœur has provided the most complete cartography: he distinguishes forgetting through the erasure of traces from reserve forgetting—what we do not remember but which remains available—and, alongside the abuses of memory (blocked, manipulated, obligated), he places the abuses of forgetting, of which amnesty—a prescribed, that is, imposed, forgetting—is the institutional form.[17] A classical precedent is Athens in 403 BC, where, after the civil war, the citizens swore *mē mnēsikakeîn*: not to recall past wrongs.[18] In 1882 Renan would make it a condition of the nation: forgetting, and even historical error, is an essential factor in its creation.[19] Pierre Nora's *lieux de mémoire*—the entities that function as emblematic referents of a community, a social group or a nation—arise precisely because there are no longer *milieux de mémoire*: they are the memory that a society fixes when it feels it is losing it.[20]

The social sciences, finally, have shown that neither remembering nor forgetting is ever purely individual. It is little more than a century since Halbwachs published *Les cadres sociaux de la mémoire* (1925): we remember within frameworks provided by the group, and when the framework disappears the memory becomes inaccessible.[21] We may add, with Jan Assmann, the distinction between communicative memory, which binds three generations, and cultural memory, which requires material supports and specialists; and, with Aleida Assmann, the difference between the canon that a society keeps active and the archive that it preserves passively.[22] The social anthropologist Paul Connerton undid the identification of forgetting with failure by distinguishing seven types of it, among them repressive erasure and the forgetting that is constitutive of a new identity.[23] I wonder how he would have

15 Plato, *Phaedrus* 274c–275b (Plato 1914). On the Aboriginal transmission of knowledge, see Yunkaporta (2019), who describes the practice of *yarning*—a narrative, non-linear conversation oriented towards learning and sharing—and Pascoe (2014).

16 Nietzsche (1874), § 1.

17 Ricœur (2000), pp. 536–589.

18 Loraux (1997). After the defeat of Athens in the Peloponnesian War, the oligarchic regime of the Thirty was established with Spartan support, which led to a civil war won by the democrats under Thrasybulus. To avoid a spiral of revenge, the reconciliation agreement of 403 BC required the citizens to swear the oath μὴ μνησικακεῖν (*mē mnēsikakeîn*), which Andocides reports as follows: "and I will harbour no grievance against any citizen, save only the Thirty, the Ten, and the Eleven: and even of them against none who shall consent to render account of his office" (Andocides, *On the Mysteries* 90, trans. Maidment, in Andocides 1941). Cf. Carawan (2012), but see the discussion of the authenticity of the documents in Canevaro and Harris (2012). The *Constitution of the Athenians* of the Aristotelian school records the clause as follows: "There should be a general amnesty concerning past events towards all persons except the Thirty, the Ten, the Eleven, and the magistrates in Piraeus; and these too should be included if they should submit their accounts in the usual way" (*Ath. Pol.* 39.6, trans. Kenyon, in Aristotle 1891).

19 Renan (1882).

20 Nora (1984–1992), especially the introduction, "Entre mémoire et histoire", in vol. I.

21 Halbwachs (1925).

22 J. Assmann (1992); A. Assmann (2016).

23 Connerton (2008). The seven types are: (1) *repressive erasure*, carried out by the State or by an authoritarian characterised the turn that AI has given to the collective memory of modern societies, given that he maintained that modernity has a particular problem with forgetting.[24] The diagnosis holds, but perhaps for reasons different from those he adduced: not so much the loss of the human scale of social life as the delegation of the selection of what is remembered (Sect. 5.3). Perhaps, then, we might add an eighth type of forgetting to his taxonomy: algorithmic forgetting, which has no agent but does have a design.

## 2.2 How the problem has been addressed: definitions and state of the art

Table 1 organises these contributions. Three ideas emerge from it, which will serve us as criteria later on.

1. Forgetting is constitutive, not deficient; a memory system that does not forget is not better but dysfunctional.
2. Forgetting is not the same as destroying. As early as 1966, Tulving and Pearlstone distinguished between the *availability* of stored information and its *accessibility* in a retrieval context; most of what we call forgetting is a loss of access, not of trace.[25] When Luciano Floridi, half a century later, analyses *Google Spain* with the same pair of concepts (Sect. 3.3), we might say that he rediscovers for the infosphere a distinction from cognitive psychology.
3. Forgetting is always social, because it depends on frameworks, material supports and collective decisions about what deserves to remain in the canon and what is left out of it.

**Table 1** Forgetting in three traditions (and in digital memory studies): notions and functions

| Field | Main notions | Function attributed to forgetting | References |
|---|---|---|---|
| Cognitive science | Decay, interference; active forgetting (inhibition); availability *vs.* accessibility; constructive memory | Adaptive: flexibility, generalisation, emotional regulation | Ebbinghaus; Bartlett; Tulving and Pearlstone; Schacter; Anderson and Hulbert; Richards and Frankland |
| Philosophy | *Lēthē*/*alētheia*; *phármakon*; active | Condition of action, of | Plato; Nietzsche; Ricœur; |

regime to eliminate inconvenient facts, images or persons from official history; (2) *prescriptive forgetting*, an act of oblivion decreed or accepted in the common interest, generally in political transitions or peace agreements, of which the Athenian amnesty is the classical example (the so-called "pact of forgetting" of the Spanish Transition is a disputed case); (3) *forgetting that is constitutive in the formation of a new identity*, the discarding of memories that would make it difficult to adopt a renewed narrative of the self; (4) *structural amnesia*, inherent in social structures such as kinship, which retain only what is relevant to the present, such as ancestors with a genealogical function; (5) *forgetting as annulment*, arising from the superabundance of information and archives, which overwhelms the capacity to process it; (6) *forgetting as planned obsolescence*, linked to the consumer economy, which rapidly discards the immediate past; and (7) *forgetting as humiliated silence*, born of shame or of a collective wound, as after a defeat or a collaboration.

24 Connerton (2009), pp. 1–6; see also Connerton (1989). Connerton was an anthropologist close to the social theory of the Frankfurt School. As he puts it: "What is being forgotten in modernity is profound, the human-scale-ness of life, the experience of living in a world of social relationships that are known. [...] the meaning of life based on shared memories" (2009, p. 5).

25 Tulving and Pearlstone (1966).

| | | | |
|---|---|---|---|
| | forgetting; reserve forgetting *vs.* erasure; prescribed forgetting (amnesty) | forgiveness and of life in common; risk of abuse | Loraux; Weinrich; Eco |
| Social sciences | Social frameworks; communicative and cultural memory; canon and archive; *lieux de mémoire*; seven types of forgetting | Construction and renewal of identities; dialectical function; selection of a usable past | Halbwachs; J. and A. Assmann; Renan; Nora; Connerton |
| Digital memory studies | Extended mind; distributed cognition; connective memory; end of decay time; algorithmic memory | Forgetting as decision and as design; forgetting through multiplication | Clark and Chalmers; Hutchins; Sparrow et al.; Hoskins; Mayer-Schönberger; Esposito; Stainforth |

Source: author's own elaboration.

## 2.3 Memory and digital environments

Cognitive processes and the human mind do not end at the boundaries of the brain, but can extend to tools and objects in the environment. The theses of the extended mind (Andy Clark and David Chalmers) and of distributed cognition (Edwin Hutchins) provide the framework for understanding what happens when memory is externalised in devices: the notebook that Otto uses to find where MoMA is in New York is part of his cognitive system, and the bridge of a ship or the cockpit of an aircraft, as Hutchins describes them, "remember" things that no crew member knows on his or her own.[26] It was later suggested experimentally—with results that a subsequent replication has called into question—that, with a search engine at hand, we remember less of the content and more of where to find it: in effect, we treat the Internet as a form of transactive memory.[27] In the terms of my earlier work, this is a case of cognitive ecology: the unit of analysis is no longer the individual but the ecosystem of agents, artefacts and rules in which cognition takes place.[28]

Memory studies have described this shift as a "connective turn" (Hoskins): from collective memory, sustained by social frameworks, to a connective memory sustained by socio-technical networks, in which the "decay time" that used to do the work of forgetting disappears.[29] Mayer-Schönberger drew normative consequences from this: if remembering has become the norm, forgetting must be reintroduced, for example through expiry dates for data.[30] This is what is now called *forgetting by design* or, as I prefer to put it, *forgetting through design*. However, as Elena Esposito has argued, it is not true that the web "never forgets": storage and access are different problems, and memory is not a store but a capacity to select. Algorithms neither remember nor forget, because they process data without understanding them; hence, whereas the challenge of human memory is to remember enough, that of algorithmic memory is to be able to forget enough, and in a controlled manner.[31] And for that very reason they can sidestep the paradox that blocked any *ars oblivionalis*—deciding to forget something obliges one to remember it—with the strategy that Umberto Eco had

26 Clark and Chalmers (1998); Hutchins (1995a, 1995b).

27 Sparrow et al. (2011). The replication by Camerer et al. (2018) did not reproduce its main result.

28 Poblet et al. (2019).

29 Hoskins (2011, 2013); Stainforth (2022), p. 258.

30 Mayer-Schönberger (2009), pp. 169–195.

31 Esposito (2017), pp. 1–2.

already pointed to: not erasing, but multiplying.[32]

Before turning to the legal and normative questions raised by this technological dimension, I must briefly introduce what we might call the “dissonance of European policy”. In 2005 Commissioner Viviane Reding, launching the digital libraries initiative that would give rise to Europeana (presented in November 2008), proclaimed that without collective memory we are nothing; in 2010, by then Vice-President of the Commission, she announced the intention to introduce a “right to be forgotten”. The first initiative rested on a static view of memory as stored heritage; the second came as a belated response to data-driven memory practices that Google had already deployed.[33] Between the duty of memory and the right to be forgotten there is no logical contradiction—they have different objects—but there is an institutional tension that the law has had to manage case by case. The sociology of traces adds a nuance: what we leave behind on the web is rarely a deliberate testimony; it consists of residues of interaction that others collect and interpret.[34]

## 3 The right to be forgotten

### 3.1 Origins and development

The right to be forgotten has three roots that should not be confused.

1. The first is *rehabilitative*: anyone who has served a sentence has the right not to be perpetually defined by his or her past. It is the root that inspires the expungement of criminal records (spent convictions), an institution whose Australian scheme my colleagues and I have formalised.[35] It is a recurrent idea in the twentieth century. It is found in the Californian decision *Melvin v. Reid* (1931); in what French legal scholarship christened the *droit à l’oubli* in the 1960s; and in the right to resocialisation that the German Federal Constitutional Court recognised in *Lebach* (1973).[36]

   It is worth noting the role of cinema and television in these origins. In *Melvin v. Reid*, the film *The Red Kimona* (1925) exploited the past of Gabrielle Darley, a former prostitute tried for, and acquitted of, the murder of her pimp; she had since left that life behind and married in 1919. In France, Claude Chabrol’s film *Landru* (1963) prompted the action brought by Fernande Segret, the serial killer’s last lover, who was still alive; the judgment of the Tribunal de Grande Instance de la Seine (1965), annotated by Gérard Lyon-Caen in 1966, lies at the doctrinal origin of the expression *droit à l’oubli*. In Germany, ZDF had prepared a docudrama on the 1969 raid on a Bundeswehr munitions depot at Lebach, in which four soldiers on guard duty were

32 Esposito (2017), pp. 6–7; Eco (1988).

33 Stainforth (2022), pp. 257–258, 261.

34 Comunello et al. (2022).

35 Governatori et al. (2020).

36 Singh and Malik (2026), p. 53 (*Melvin v. Reid*); Weber (2011), p. 121 (*Lebach*); on the French and Italian antecedents, Mantelero (2013).

murdered; one of those convicted, sentenced to six years' imprisonment as an accessory, was about to be released on parole.[37]

2. The second root is *data protection*: the principles of purpose limitation and storage limitation, and the rights to erasure and to object laid down in Directive 95/46/EC (Arts. 12 and 14), which speak not of forgetting but of the processing of data.
3. The third lies in the protection of *dignity*: personality rights—honour, privacy, one's own image—and, earlier still, the "right to be let alone", an expression coined by Judge Thomas M. Cooley that Samuel Warren and Louis Brandeis popularised in "The Right to Privacy" (*Harvard Law Review*, 1890).[38]

Robert Post has insisted that the latter two approaches follow different logics. Data privacy is instrumental and bureaucratic. The dignitary function, by contrast, protects social norms of respect. His thesis is that much of the subsequent confusion stems from having merged them.[39]

The contemporary formulation is European. The 2012 proposal for a General Data Protection Regulation included a "right to be forgotten and to erasure"; the final text (Regulation (EU) 2016/679, GDPR) enshrines it in Article 17 as the "right to erasure ('right to be forgotten')", with six enabling grounds and five exceptions, among them freedom of expression and information, and archiving purposes in the public interest or scientific or historical research purposes.[40] Article 17(2) adds the obligation, for a controller who has made the data public, to inform other controllers of the request: an early recognition that information circulates in chains. Spanish Organic Law 3/2018 splits the right into two modalities—Internet searches (Art. 93) and social networks and equivalent services (Art. 94)—with reinforced protection for minors. Outside Europe the picture is fragmentary: Alnimer sees a "fragmented world" in which the European extraterritorial claim collides with diverse digital sovereignties, and proposes, following Svantesson's layered approach, a multilevel system.[41]

### 3.2 European and national case law

Table 2 sets out the main decisions. Two lines of case law intersect. The Court of Justice of the European Union (CJEU) has built, on the basis of data protection, a right to delisting (de-indexing) vis-à-vis search engines. The European Court of Human Rights (ECtHR), on the basis of the balance between Articles 8 and 10 of the European Convention on Human Rights, has had to decide whether forgetting can reach the source: the digital press archive.

*Google Spain* (2014) has its origins close to home. In 1998 *La Vanguardia* had published, by legal mandate, two notices of real-estate auctions for social security debts in which Mario

37 *Melvin v. Reid*, 112 Cal. App. 285 (1931); TGI Seine (1965), *JCP* 1966 II 14482, note by G. Lyon-Caen (the court dismissed the action because Segret herself had published her memoirs); BVerfGE 35, 202 (*Lebach*).

38 Cooley (1888), p. 29; Warren and Brandeis (1890), p. 195.

39 Post (2018), pp. 981–983.

40 Regulation (EU) 2016/679, Art. 17 and recitals 65–66.

41 Alnimer (2026), pp. 1–2; see Svantesson (2013).

Costeja's name appeared. Once the newspaper's archive had been digitised, a search for his name returned them in first place a decade later, although the debt had been settled. The Spanish Data Protection Agency rejected the complaint against the newspaper and upheld the one directed against Google; the Audiencia Nacional referred a question for a preliminary ruling. The Court held that the operator of a search engine processes personal data and is the controller of that processing; that its activity affects fundamental rights in a way that is additional to, and distinct from, that of the publisher, because it makes it possible to obtain "a structured overview" of the person; and that the data subject may request the removal of links to lawful and truthful information when, owing to the time that has elapsed, it has become inadequate, irrelevant or excessive, unless the person's role in public life justifies a preponderant interest of the public.[42] Its most lasting contribution is procedural: separating the survival of the source from its retrievability through a search by name.[43]

**Table 2** Main decisions on the right to be forgotten

| Court and date | Case | Doctrine |
|---|---|---|
| CJEU, 13.5.2014 | C-131/12, *Google Spain* (Costeja) | The search engine is a controller; right to the delisting of lawful information that has become inadequate, irrelevant or excessive with the passage of time; limited by the data subject's role in public life. |
| CJEU, 9.3.2017 | C-398/15, *Manni* | In principle, no right to be forgotten in respect of the companies register; exceptionally, limited access after a sufficiently long period. |
| CJEU, 24.9.2019 | C-136/17, *GC and Others v. CNIL* | Special categories of data and data relating to criminal convictions also bind the search engine; the ranking of results must reflect the current judicial situation. |
| CJEU, 24.9.2019 | C-507/17, *Google v. CNIL* | Territorial scope: delisting on the versions for the Member States, with measures discouraging access from within the EU; EU law neither requires nor prohibits worldwide delisting. |
| CJEU, 8.12.2022 | C-460/20, *TU and RE v. Google* | Inaccurate information: delisting if the applicant provides relevant and sufficient evidence of manifest inaccuracy, without the need for a prior judicial decision; extends to thumbnails. |
| ECtHR, 16.7.2013 | *Węgrzynowski and Smolczewski v. Poland* | It is not for the courts to rewrite history by ordering the removal of an article from the digital archive; an explanatory note is preferable. |
| ECtHR, 28.6.2018 | *M.L. and W.W. v. Germany* | Refusing the anonymisation of press archives concerning a notorious murder does not violate Art. 8: persistent public interest and the role of archives. |
| ECtHR, 25.11.2021 | *Biancardi v. Italy* | A publisher's liability for failing to delist an article does not violate Art. 10; the duty does not fall on search engines alone. |
| ECtHR (GC), 4.7.2023 | *Hurbain v. Belgium* | Ordering the anonymisation of *Le Soir*'s digital archive does not violate Art. 10; seven criteria: nature of the information, time, current interest, notoriety, repercussions, accessibility, impact of the measure. |
| BVerfG, 5.6.1973; 6.11.2019 | *Lebach*; *Recht auf Vergessen I* and *II* | Right to resocialisation; time as a balancing factor; graduated obligations of press archives; interplay of national fundamental rights and the Charter. |
| Spanish Supreme Court (Civil Chamber, Plenary), 15.10.2015; Spanish | STS 545/2015; STC 58/2018 (*El País* digital press archive) | The publisher must prevent indexing by external search engines; the Constitutional Court adds the internal search engine by name; the archive remains unaltered. |

42 CJEU (Grand Chamber), C-131/12, *Google Spain*, paras. 28–41, 80–81 and 93–99.

43 Singh and Malik (2026), p. 52; Frantziou (2014).

| | | |
|---|---|---|
| Constitutional Court, 4.6.2018 | | |
| Cass. (Sez. Un.), 22.7.2019 | Judgment no. 19681 | Recalling past events is historiographical activity; mentioning the name requires a current public interest. |

Source: author's own elaboration based on the decisions and on Kulk and Zuiderveen Borgesius (2018), Askar (2026) and Singh and Malik (2026).

Subsequent decisions have delimited the right on three fronts. *By object*: in principle it does not cover public registers (*Manni*), but it does cover sensitive and criminal data (*GC*), as well as manifestly inaccurate information, with a reasonable burden of proof on the applicant (*TU and RE*). *By territory*: delisting is, as a rule, European and not worldwide (*Google v. CNIL*). And *by source*: Strasbourg has moved from a reluctance to "rewrite history" (*Węgrzynowski*) to accepting, in *Hurbain*, that the anonymisation of a digital press archive may be a proportionate interference, while laying down a seven-criterion test.[44] National courts have explored intermediate solutions: in Spain, delisting by the publisher without touching the archive; in Germany, obligations graduated according to time; in Italy, the distinction between news reporting and historiography. The result is less a unitary right than a *ladder of remedies*—erasure, delisting, masking or anonymisation, removal, access restriction—from which the least restrictive effective option must be chosen. This direction can also be observed outside Europe, for example in the judgment of the High Court of Delhi of 29 May 2026 (*Laksh Vir Singh Yadav v. Union of India*), which distinguishes between preservation, masking and name-based delisting in relation to court decisions published online.[45]

### 3.3 Paradoxes, controversies and debates

Few legal institutions have provoked so lively a debate. In its favour it has been argued (i) that the right restores a minimum of individual control over one's informational identity and gives effect to the consent regime (it has been proposed to limit it precisely to processing based on consent); (ii) that it protects the possibility of reinventing oneself and keeps the future open; and (iii) that it corrects a power asymmetry between persons and platforms.[46] Against it, critics saw in it—even before the judgment—(i) "the biggest threat to free speech on the Internet in the coming decade" and (ii) a head-on clash with the American tradition of protecting truthful information.[47] Post adds that *Google Spain* applies the bureaucratic logic of data privacy to an actor, the search engine, that sustains the contemporary public sphere just

44 ECtHR (Grand Chamber), *Hurbain v. Belgium*, para. 205; Singh and Malik (2026), pp. 52–53. The seven criteria are: (1) the nature of the archived information; (2) the time elapsed since the events and since the initial and online publication; (3) the current interest of the information; (4) the notoriety of the person and his or her conduct since the events; (5) the negative repercussions of the continued availability of the information online, such as the risk of a perpetual virtual criminal record; (6) the degree of accessibility of the information in the digital archive; and (7) the impact of the measure on freedom of expression and, in particular, on freedom of the press. On the domestic origin of the case (Belgian Cour de cassation, 29 April 2016) and on *Biancardi*, see Askar (2026), pp. 457–458.

45 High Court of Delhi, *Laksh Vir Singh Yadav v. Union of India*, W.P.(C) 1021/2016, 29 May 2026, according to Singh and Malik (2026), pp. 50–51, 55.

46 Ausloos (2012); Weber (2011).

47 Rosen (2012), p. 88; Fazlioglu (2013), p. 149.

as the press sustained that of the nineteenth century, and that the Court ought to have reasoned from dignitary privacy, which is capable of balancing.[48]

Luciano Floridi, one of the eight independent members of the advisory council that Google set up after the judgment, has offered a philosophical reading of the judgment. The key is the distinction between availability and accessibility: the information remains published, but ceases to be linked. In an environment in which we access content through links, removing the link amounts to hiding the content; in the infosphere, he says, the map becomes more important than the territory.[49] It is, as we have seen, Tulving and Pearlstone's distinction applied to social memory. And it explains why the solution is at once modest and effective: it erases nothing, but it reintroduces the constraint that time no longer provides.

The debate has left at least six well-identified paradoxes, to which I add a seventh:

- *The Costeja paradox*. The citizen who wished to be forgotten is today the proper name of a judgment studied in every law faculty. Floridi observes that, by the Court's own criteria, one could argue that he has become a public figure and that the information is now relevant.[50]
- *Remembering in order to forget*. Every decision to forget points to what it seeks to conceal. From delisted articles and public data, a third party can rediscover some of the removed links and identify some of the requesters.[51] A variant of the so-called "Streisand effect"—named after the lawsuit that Barbra Streisand brought in 2003 to have a photograph of her house removed, which only increased its circulation—is at work here: attempting to conceal a matter can arouse curiosity and multiply its impact. It is a real possibility, though not a generalised one.
- *The privatisation of judgment*. The balancing of privacy against the public interest is carried out in the first instance by a private company, according to its own criteria and without any obligation to hear the publisher. Google's own study of five years of application—3.2 million URLs requested; a thousand requesters, 0.2% of the total, responsible for 16% of the requests—illustrates the scale and opacity of the process.[52]
- *The misleading name*. Since no one can be obliged to forget, the right does not guarantee forgetting: it produces, at best, practical obscurity. The name promises more than it delivers and feeds both the expectations of right-holders and the alarm of critics.
- *Territory*. A delisting limited to Europe can be circumvented; a worldwide one exports the European balance between privacy and expression to third countries.
- *The individual and the archive*. Each person's forgetting affects the memory of others: that of those involved in the same events, that of future historians, that of the

48 Post (2018), pp. 981–983.

49 Floridi (2015), pp. 167–168.

50 Floridi (2015), p. 177.

51 Xue et al. (2016), pp. 395–397, 401.

52 Bertram et al. (2019), p. 959; see also Kulk and Zuiderveen Borgesius (2018).

community. The individual right is exercised over a good that is, in part, common.[53]

– *The de-identification paradox*. Sensitive data are published "de-identified" precisely in order to protect them; but de-identification can fail, and whoever demonstrates this may run up against the reaction of those in power, because the matter is immediately politicised and nobody likes to be contradicted or exposed. This happened in Australia in 2016–2017. The Department of Health openly published a "de-identified" sample of billing data from the Medicare Benefits Schedule (MBS) and the Pharmaceutical Benefits Scheme (PBS) covering 10% of the population. Chris Culnane, Benjamin Rubinstein and Vanessa Teague showed first that the encrypted provider identifiers could be reversed, and then that patients could be re-identified without decrypting anything, by cross-matching unencrypted fields—dates of medical procedures, year of birth—with external information.[54] Removing direct identifiers (names, Medicare numbers) is insufficient if quasi-identifiers make it possible to build unique profiles. The Government withdrew the dataset and proposed making re-identification a criminal offence, which triggered an important debate on the publication of open data. It is a case I witnessed at close quarters: the researchers, who achieved this in very little time and reported it to the Department in the belief that they were helping, came under pressure from the Department of Health itself, as did their university. De-identification is the institutional form of data forgetting, and the case shows that this technical forgetting can fail.

## 4 State of the art in the specialised literature (2009–2026)

If we take as our threshold the publication of *Delete* (2009) and the announcement of the European reform (2010), the literature can be arranged in four periods (Table 3). They are not watertight compartments, but each responds to a main question, displays distinctive features and is addressed from several disciplines.

**Table 3** Periodisation of the literature on the right to be forgotten

| Period | Main question | Features | Representative works |
|---|---|---|---|
| I. Conceptual and programmatic (2009–2014) | Should it exist? What is it? | Transnational normative debate; genealogy (*oubli*, *oblio*, *Lebach*); technical proposals (expiry dates) | Mayer-Schönberger (2009); Koops (2011); Weber (2011); Rosen (2012); Ausloos (2012); de Terwangne (2012); Simón Castellano (2012); Mantelero (2013); Fazlioglu (2013) |
| II. Jurisprudential and balancing (2014–2018) | How is it balanced? Who decides? | Commentary on *Google Spain*; search engine as controller; criteria (time, public role); first empirical studies | Frantziou (2014); Pagallo and Durante (2014); Floridi (2015); Sartor (2016); Jones (2016); Xue et al. (2016); Esposito (2017); Kulk and Zuiderveen Borgesius (2018); Post (2018) |
| III. Implementation and scope (2018– | How far does it reach? Does it | GDPR Art. 17; territoriality; press archives; transparency; | Fosch-Villaronga et al. (2018); Veale et al. (2018); Bertram et al. (2019); Bourtoule et |

53 Esposito (2017), p. 3; Stainforth (2022), p. 265.

54 Culnane et al. (2017). The legislative proposal was the Privacy Amendment (Re-identification Offence) Bill 2016.

| | | | |
|---|---|---|---|
| 2022) | work? | collective memory; first warnings about machine learning | al. (2021); Stainforth (2022); Comunello et al. (2022) |
| IV. Generative (2022–2026) | Can an AI forget? What does forgetting mean? | Language models, hallucinations, *deepfakes*; unlearning; GDPR and AI Act; new subjects (children, the deceased) and geographies; algorithmic memory | Hoskins (2024); Cooper et al. (2025); Zhang et al. (2025); Tan (2025); Tutuncuoglu (2025); Reddy et al. (2025); Askar (2026); Fernández Delgado and Cazurro Barahona (2026); Michałkiewicz-Kądziela and Staszewski (2026); Cheriti et al. (2026); Alnimer (2026); Singh and Malik (2026) |

Source: author’s own elaboration. The periods overlap in the transition years.

The first period is *conceptual*. The discussion concerns whether the right should exist, along a marked transnational axis, and the dominant discipline is constitutional and information law.[55] The second is *jurisprudential*: the *Google Spain* judgment shifts the question from *whether* to *how*, and the philosophy of information (Floridi), legal theory (Sartor, with his analysis of the evolution of interests in the flux of time; Pagallo and Durante) and sociology (Esposito) join the debate.[56] The third is one of *implementation*: with the GDPR applicable from 2018, the literature becomes empirical and comparative, and the first works appear that warn that machine-learning systems “remember” in a way the law had not foreseen—humans forget, machines remember—together with the demonstration that models can be subjected to inversion attacks that recover training data from them.[57]

The fourth period, in which we now find ourselves, is *generative*. The public release of conversational large language models at the end of 2022 changes the object. Reddy and colleagues document bibliometrically the growth of *machine unlearning* as a field, explicitly motivated by the GDPR’s right to be forgotten.[58] Three shifts can be observed:

(i) *Disciplinary*: the centre of gravity moves from constitutional law to computer science, with the risk that “forgetting” is redefined as whatever is technically measurable.

(ii) *Right-holders*: the right extends to holders who cannot exercise it themselves: for example, children who are victims of digital sexual violence, for whom the slowness of the remedy is a form of revictimisation (Irwan et al.), or deceased persons, whose digital remains now feed AI-generated content (Al-Mutairi).[59]

(iii) *Geographical*: the 2025 and 2026 contributions I have been able to consult include work from the Gulf, India, Indonesia, China, Poland, Spain and Algeria. The right to be forgotten has ceased to be a European singularity and has become a common reference point of global digital constitutionalism, with all the divergences this entails.

55 De Terwangne (2012); Simón Castellano (2012). In the Spanish and Catalan context, the works of Cécile de Terwangne in the UOC journal *IDP* and of Pere Simón Castellano soon set the terms of the debate.

56 Sartor (2016); Pagallo and Durante (2014).

57 Fosch-Villaronga et al. (2018); Veale et al. (2018).

58 Reddy et al. (2025).

59 Irwan et al. (2026), p. 66; Al-Mutairi (2026).

## 5 The right to be forgotten in contact with artificial intelligence

### 5.1 The third regime: from the locatable datum to the distributed representation

Figure 1 summarises the argument. In the regime of the *archive*, memory resides in documents; the operation is preservation, and the remedy is erasure or anonymisation. In the regime of the *search engine*, the relevant memory is the index; the operation is retrieval and ranking, and the remedy is delisting. In both cases there is a discrete object—a record, a link—and an identifiable controller. In the regime of the *generative model*, learnt information is not stored as a record but distributed across billions of parameters; the operation is inference and generation; and there is no "place" to which one can go to find the datum in order to erase it. To the availability/accessibility pair a third term must be added, which I propose to call *generability*: the capacity of a system to produce information about a person, whether or not that information is present in any source.

| | **I. ARCHIVE**<br>*Availability · 1995–2014* | **II. SEARCH ENGINE**<br>*Accessibility · 2014–2022* | **III. GENERATIVE MODEL**<br>*Generability · 2022–* |
|---|---|---|---|
| **Where memory resides** | Document, record, press archive | Index, ranking, aggregated profile | Parameters, vectors, agent memory |
| **Operation** | Store and preserve | Retrieve and rank | Infer and (re)generate |
| **Forgetting remedy** | Erasure, anonymisation, cancellation | Delisting, practical obscurity | Unlearning, filters, control over identity |
| **Limit of the remedy** | Historical interest, freedom of information | Territoriality, Streisand effect | Non-locatability, verification, hallucination |

*From the discrete record to distributed representation: forgetting ceases to be a local operation*

**Fig. 1** Three regimes of digital memory and their forgetting remedies. Source: author's own elaboration

That this poses a legal, and not merely a technical, problem was debated for a time. In July 2024 the Hamburg data protection authority maintained that the parameters of a model are not personal data, because they do not store information about specific individuals. The European Data Protection Board, in Opinion 28/2024 (December 2024), adopted a more demanding position: a model trained on personal data cannot be considered anonymous unless it is shown that the likelihood of extracting such data from it, or of obtaining them through queries, is insignificant.[60] Technical research supports this view: verbatim fragments of training data, including personal information, can be extracted from language models.[61]

The case that best illustrates the mismatch is the complaint that *noyb* (*None of your business* – European Center for Digital Rights), Max Schrems's association, lodged against OpenAI with the Austrian data protection authority in 2024. The system attributed an

60 HmbBfDI (2024); EDPB (2024); Fernández Delgado and Cazurro Barahona (2026), p. 3.
61 Carlini et al. (2021).

incorrect date of birth to a public figure; when asked to rectify or erase it, the company acknowledged that it could not prevent the model from displaying the wrong date without blocking all information about the data subject: the filter could only be applied globally, not selectively.[62] A second complaint, lodged in 2025 with the Norwegian authority, concerned a citizen to whom the system falsely attributed the murder of two of his children. Fernández Delgado and Cazurro Barahona conclude that we are not facing a regulatory gap but a compliance problem, and that the solutions available are, for the time being, palliative rather than structural.[63]

What might the solution be? Machine unlearning is the technical answer. The idea is to remove the influence of certain training examples without retraining the entire model.[64] Research has pointed out its limits with a clarity that lawyers should bear in mind: removing information from the training set does not guarantee that the model cannot generate it, because it may infer it from other data; suppressing outputs is not erasing; and legal objectives cannot be translated without remainder into technical metrics.[65] From the legal side, several proposals have been made: to interpret the concept of "erasure" broadly (Tan) or, noting that the AI Act provides no specific forgetting mechanism for generative systems, to create mechanisms of algorithmic oversight, such as a dedicated committee (Askar).[66] I would add one observation. The parallel with cognitive science (Sect. 2.1) is instructive but misleading. Models already "forget" in Richards and Frankland's sense—they generalise, regularise, lose details; what they do not do is forget at the will of a third party and in a verifiable manner. And the right to be forgotten has never consisted in the forgetting of the system, but in a person's power over what the system does with him or her.

### 5.2 From remembering to generating: identity, hallucinations and *deepfakes*

The second feature of the new regime is that harm no longer stems only from the persistence of truthful information. *Google Spain* concerned true facts that time had rendered irrelevant. Generative systems produce statements about persons that have never been true—hallucinations—and images, voices and videos of events that never occurred. In the face of *deepfakes*, Article 17 loses effectiveness: it presupposes a pre-existing datum that can be identified and erased, whereas synthetic content is created and replicated faster than any procedure. Michałkiewicz-Kądziela and Staszewski therefore propose redefining the right as a right to control one's own digital identity, with preventive *ex ante* mechanisms and not merely reactive ones.[67]

The proposal is debatable in its details, but its diagnosis is sound: the object of the right has shifted from information *about* me to the capacity to *represent* me. This connects the right to be forgotten with the dignitary root that Post championed, and with two limiting cases identified in recent literature. The first is that of the deceased, whose traces feed AI-generated content—and, increasingly, conversational avatars: who exercises forgetting on behalf of those who are no longer here, and against whom?[68] The second is that of neurodata, where what is inferred is not what we have done but what we think or feel.[69]

---

62 Fernández Delgado and Cazurro Barahona (2026), p. 8; noyb (2024).

63 Fernández Delgado and Cazurro Barahona (2026), pp. 8–9. On the second complaint, lodged with the Norwegian authority (Datatilsynet), see noyb (2025).

64 Cao and Yang (2015); Bourtoule et al. (2021).

65 Cooper et al. (2025).

66 Tan (2025), p. 60; Askar (2026), pp. 451, 469–472.

67 Michałkiewicz-Kądziela and Staszewski (2026), pp. 161–163.

### 5.3 The prefiguration of collective memory

So far, the individual problem. But AI does not merely make it harder for someone to be forgotten; it reconfigures what a society remembers. *Algorithmic memory* is the name given to the process by which AI systems, in deciding what is stored, what surfaces and what is erased, do not merely process information but construct memory at three levels: personal, institutional and social.[70] Cheriti and colleagues describe AI as a "silent manufacturer" of collective memory and an "invisible editor" of media history: algorithms that suppress politically sensitive material, search engines that obscure commercially inconvenient content, automated tools that privilege some archival materials over others.[71] Forgetting, here, is not a right that is exercised but an effect that is suffered. Connerton would have recognised in it features of repressive erasure, but without any identifiable censor; this forgetting comes closer to what he called structural amnesia, or to forgetting as annulment arising from superabundance—even if, as I have suggested, it deserves a type of its own.

Three mechanisms seem to me particularly relevant.

1. The first is *selection*: the hierarchy of results, recommendations and summaries acts as a canon, in Aleida Assmann's sense, and relegates the rest to an archive that nobody consults.
2. The second is *regeneration*. Spennemann asked a language model (ChatGPT-4o) to imagine humanity's cultural heritage thirty years from now, in an exercise that is a contemporary equivalent of the Côté postcards we discussed last year; and the review by Sui and colleagues shows that the generative restoration of heritage—completing the fresco, reconstructing the monument—is already a common practice that raises problems of cultural coherence and authenticity.[72] The past we shall see will be, in part, a synthesised past.
3. The third is *feedback*. Recent experiments indicate that a conversational agent based on language models induces significantly more false memories than a questionnaire, and that AI-edited images and videos can implant memories of events never experienced: Schacter's suggestibility and misattribution, industrialised.[73] And at the collective scale, Shumailov and colleagues have shown that models trained

68 Al-Mutairi (2026).

69 Fernández Delgado and Cazurro Barahona (2026), pp. 4, 7.

70 Tutuncuoglu (2025), p. 1.

71 Cheriti et al. (2026), pp. 26, 32.

72 Spennemann (2026); Sui et al. (2026), pp. 437, 443–444.

73 Chan et al. (2024); Pataranutaporn et al. (2025).

> recursively on data generated by other models collapse: they first forget the tails of the distribution, that is, the infrequent, the marginal, the singular.[74] For a medium-sized language and culture such as Catalan, this is not a technical curiosity. It is a form of structural forgetting that requires no one's decision.

This is where the thread connecting this lecture to last year's becomes visible. Pedreschi and colleagues describe human–AI coevolution as a loop in which users' decisions generate data that train models which, in turn, shape users' subsequent preferences.[75] Replace "preferences" with "memories" and you have the prefiguration of memory. Among the risks of the hybrid society identified in last year's lecture were four forms of relaxation: practical, epistemic, ethical and symbolic. A fifth should now be added, *mnemonic relaxation*: the delegation to machines not of remembering as such—we have done that since the invention of writing—but of the selection of what is worth remembering and forgetting. The digital *Deixis am Phantasma* of which I spoke last year, the possibility of entering our imagination and dwelling within it, also holds for the past: we may end up inhabiting a memory that we have not formed ourselves. This has been documented for diasporic identities, where digital narratives can both preserve and undermine collective memory.[76]

### 5.4 From the individual solution to the governance of the ecosystem

If the analysis is correct, the right to be forgotten cannot be guaranteed in the third regime with the instruments of the first two: a request, a controller, a balancing exercise, an act of erasure or delisting. Not because these instruments are mistaken—they remain necessary for archives and search engines, and for the retrieval layer of AI systems—but because they are insufficient. In earlier work I have defined *legal governance* as the set of processes that generate a sustainable regulatory ecosystem reflecting the fundamental concepts of a democracy; *legal ecosystems* as complex and dynamic systems with multiple levels and actors; and *smart legal ecosystems* as regulatory systems partially embedded in cyber-physical systems, oriented towards real-time compliance.[77] The model distinguishes four normative clusters—hard law, policies, soft law and ethics—and places validity at the centre.[78] The approach is neither top-down nor bottom-up, but *middle-out* and *inside-out*: rules are articulated in the very design of the systems.

74 Shumailov et al. (2024).

75 Pedreschi et al. (2025). On the four forms of relaxation, see Casanovas (2026), Sect. 5.

76 Kianpour et al. (2026), pp. 36, 42.

77 Casanovas and Noriega (2026); Casanovas et al. (2022, 2024).

78 Casanovas (2025d); Casanovas et al. (2026).

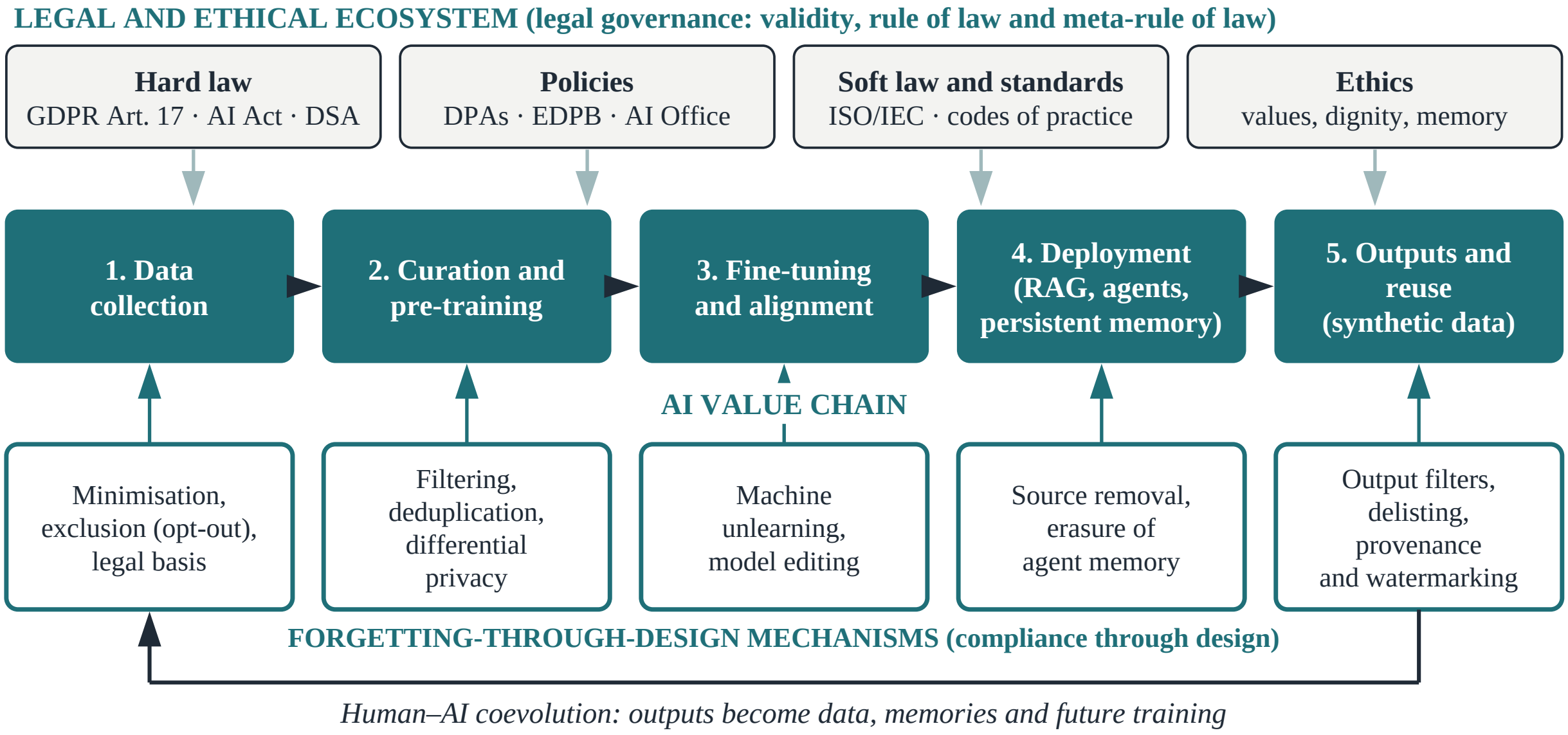

**Fig. 2** The right to be forgotten distributed along the AI value chain within a legal and ethical ecosystem. Source: author's own elaboration based on Casanovas (2024, 2025f) and Casanovas and Noriega (2026)

Applied to our problem (Fig. 2), this means three things. First: forgetting must be *distributed* along the AI value chain. Each link has a different mechanism—minimisation and exclusion at collection; filtering, deduplication and differential privacy in training; unlearning and editing in fine-tuning; removal of sources and deletion of persistent memories in deployment; filters, delisting and provenance in outputs—and a different responsible party. I have argued elsewhere that the value chain, an economic concept that the European legislator has turned into a legal one, must incorporate the intangible dimensions—language, culture, ethical and legal values—that are not directly monetisable;[79] memory is one of them. Second: forgetting must be guaranteed *through design*, as a particular case of compliance through design, and it must be possible to *validate* it empirically, not merely to declare it. Third: what is specifically legal about this right must be preserved. Balancing a person's dignity against a community's memory cannot be automated; what can be designed is the environment in which this balancing is carried out with safeguards, with traceability and with a human being who is accountable for it.[80] This needs to be said now, when the European debate on regulatory simplification could weaken the safeguards attached to the legal bases for processing personal data to train models: the false choice between regulation and innovation is paid for, among other currencies, in the capacity to forget.[81]

## 6 Conclusions: what can we expect?

I summarise the argument in six theses and three expectations.

1. *Forgetting is a function, not an interruption of the system.* Cognitive science, philosophy and the social sciences converge in regarding it as constitutive of

79 Casanovas (2025f); see also Casanovas (2025a).

80 Casanovas et al. (2023).

81 Casanovas (2025e).

individual and collective memory. An environment that does not forget is not more faithful; it is dysfunctional.

2. *The right to be forgotten is a ladder of remedies, not a unitary right.* It has three roots—rehabilitative, data-protective and dignitary—and the case law has given it concrete form in erasure, delisting, masking or anonymisation, removal and access restriction, to be chosen according to proportionality. The name promises more than it delivers; what it does deliver—constraint, practical obscurity—is valuable.
3. *It was conceived for two regimes of memory, the archive and the search engine, and now operates in a third.* In the archive and in the search engine there is a locatable object and a controller. In the generative model, information is distributed, inferred and regenerable. To availability and accessibility, generability is added.
4. *The object of the right is shifting from information about the person to the capacity to represent him or her.* Hallucinations, *deepfakes*, posthumous avatars and neurodata point towards a right to control one's digital identity, which recovers the dignitary root.
5. *AI prefigures collective memory through selection, regeneration and feedback.* The structural forgetting of the tails of the distribution—the marginal and the singular—is a specific risk for languages and cultures such as Catalan. The individual right to be forgotten and the collective duty of memory are not contradictory: both stand against a memory administered without anyone answerable for it.
6. *The effective exercise of the right requires moving from the individual* ex post *solution to the governance of the ecosystem*: forgetting distributed along the value chain, guaranteed through design, open to empirical validation, and with a human, public forum for balancing.

What can we expect? In the short term, that the European courts will rule on the application of Article 17 to generative models, probably extending the logic of *Google Spain*: whoever determines the purposes and means is answerable, and technical difficulty is no exemption. In the medium term, an imperfect convergence between law and technology: unlearning will never fully satisfy the ideal of erasure, just as delisting did not satisfy that of forgetting, but it may come to provide sufficient obscurity if accompanied by verification mechanisms. In the long term, the question is political: who decides what a hybrid society remembers and forgets? The worst answer is that no one decides, because then de facto powers, both political and corporate, can occupy that place and decide for us, without having to answer for it.[82]

Last year I closed with the image of the giant waves that no sailor who has seen them has survived to describe: we have to imagine them. Forgetting resembles them. By definition we cannot see it; we perceive only its edges, its blurred limits, when we look for something and it is not there, or when we find what ought to have disappeared. Junceda had his characters of 1875, 1925 and 1975 exclaim: "Ah! Oh! A car!", "Ah! Oh! A horse!", "Ah! Oh! Someone on foot!" It is not hard to imagine the next panel. The year 2075: "Ah! Oh! Someone who

82 On these risks, see Casanovas (2025b, 2025c).

forgets!” Whether this turns out to be a rarity or a right will depend on what we design now.

## Acknowledgements

This work was carried out in part within the framework of the activities of the UAB associated unit at IIIA-CSIC, Research Group on AI and Law (2023–2026), and of the CSIC project CAI – Civic Artificial Intelligence: Artificial Intelligence at the Service of Society (ILINK25174). In preparing this text I have used Claude (Anthropic) as an assistant for the bibliography, source verification, editing and the preparation of tables and diagrams, and I have also taken up some of its suggestions. Responsibility for the content is mine alone.

## Case law and legislation

# L'oblit prefigurat

## Memòria, dret a l'oblit i intel·ligència artificial

**POMPEU CASANOVAS**

*Institut de Recerca en Intel·ligència Artificial (IIIA-CSIC)*
*La Trobe University*
*Institut de Dret i Tecnologia (IDT-UAB)*
*pompeu.casanovas@iiia.csic.es · https://orcid.org/0000-0002-0980-2371*

**Resum.** Aquesta lliçó inaugural estableix l'estat de l'art del dret a l'oblit (*right to be forgotten*) i n'examina la transformació en contacte amb la intel·ligència artificial. Parteix de la ciència cognitiva, la filosofia i les ciències socials per mostrar que l'oblit no és un dèficit de la memòria sinó una de les seves funcions constitutives, i que recordar el passat i imaginar el futur comparteixen mecanisme. Reconstrueix després els orígens, la jurisprudència europea i nacional, i les paradoxes del dret a l'oblit, i proposa una periodització de la literatura especialitzada (2009–2026). La tesi central és que la memòria digital ha travessat tres règims —arxiu, cercador i model generatiu— i que el dret fou concebut per als dos primers. En el tercer, la informació personal ja no es conserva com a registre localitzable sinó que s'infereix i es regenera; la IA prefigura els marcs del record individual i col·lectiu. El dret a l'oblit deixa de ser només un recurs individual *ex post* i esdevé una propietat que cal dissenyar i governar en ecosistemes jurídics i ètics al llarg de la cadena de valor de la IA.

**Abstract.** This inaugural keynote lecture surveys the state of the art of the right to be forgotten and examines how it is transformed by artificial intelligence. Drawing on cognitive science, philosophy and the social sciences, it treats forgetting as a constitutive function of memory, and recalls that remembering the past and imagining the future share the same mechanism. It then reviews the origins, the European and national case law, and the paradoxes of the right, and offers a periodisation of the literature (2009–2026). The core claim is that digital memory has moved through three regimes —archive, search engine and generative model—and that the right was designed for the first two. In the third, personal information is no longer stored as a locatable record but inferred and regenerated; AI prefigures the frameworks of individual and collective remembering. The right to be forgotten ceases to be merely an individual *ex post* remedy and becomes a property to be designed and governed within legal and ethical ecosystems along the AI value chain.

### 1 Introducció: del futur imaginat a l'oblit prefigurat

L'any passat ens vam preguntar en aquests Col·loquis com representem el futur. Vaig defensar aleshores que les imatges de l'esdevenidor —les de Junceda a *En Patufet* de 1926, les postals de Côté, les targetes d'Echte Wagner— són portadores de valors, tenen un subjecte col·lectiu i estan ancorades en el món quotidià de qui les produeix. El límit de la nostra capacitat de prefigurar, deia, és que «ens guiem per allò que podem veure», i que les nostres percepcions estan limitades per fantasmes simbòlics que tenim a tocar.[1] Enguany el tema és l'oblit. Sembla un gir de cent vuitanta graus —del demà a l'ahir—, però voldria mostrar que és el mateix problema vist des de l'altre costat.

Ho és, en primer lloc, per una raó cognitiva. La recerca sobre memòria episòdica ha establert que recordar el passat i imaginar el futur usen en bona part el mateix sistema: la memòria no és un arxiu sinó un procés constructiu que recombina fragments d'experiència, i és precisament aquesta flexibilitat la que ens permet simular escenaris que encara no han passat.[2] Karl Bühler ho havia vist des del llenguatge: la *Deixis am Phantasma*, el tercer mode de la dèixi al costat de la *demonstratio ad oculos* i de l'anàfora, assenyala per igual allò recordat i allò fantasiejat, perquè en tots dos casos el parlant es desplaça a un camp que no té al davant.[3] Si això és així, qui intervé en la matèria primera dels records intervé també en el repertori del futur imaginable. El que una societat pot esperar depèn del que pot recordar, i del que se li permet d'oblidar.

Ho és, en segon lloc, per una raó biogràfica que em permeto d'esmentar. El 2007 els Col·loquis es van dedicar a la memòria, i hi vaig presentar un treball sobre memòria escrita i imatge gràfica.[4] Dinou anys després, el cicle es tanca amb el seu revers. Entremig ha passat una cosa que aleshores no podíem preveure: la condició per defecte s'ha invertit. Durant mil·lennis, oblidar era el que succeïa si no es feia res, i recordar exigia esforç, tècnica i institucions. Avui conservar és gratuït i automàtic, i és l'oblit el que demana una decisió, un cost i —aquest és el nostre tema— un dret.[5]

El dret a l'oblit és la resposta jurídica europea a aquesta inversió. Va néixer en un entorn de documents i arxius, es va consolidar davant dels cercadors i es troba ara amb una tecnologia que no encaixa en cap dels dos motlles. Els models d'intel·ligència artificial generativa, deia l'any passat, introdueixen la probabilitat «com a element de fet en la nostra forma de sentir, pensar, argumentar i decidir».[6] Ara hi hem d'afegir: també en la nostra forma de recordar i de ser recordats. Un model de llenguatge no conserva una notícia sobre mi; n'ha après regularitats estadístiques i, quan algú li pregunta, en genera una de versemblant. No evoca: prefigura.

Faig servir aquest verb deliberadament. Prefigurar és configurar per endavant: disposar l'entorn, el context i els marcs dins dels quals alguna cosa podrà aparèixer. La meva tesi és que la IA té avui la capacitat de prefigurar la memòria —individual i col·lectiva— en tres sentits: *selecciona* el que es fa visible del passat, *regenera* el que ja no hi és (o no hi ha estat mai) i *retroalimenta* amb els seus productes els records humans i les dades dels models futurs. I que això afecta la possibilitat d'exercir efectivament el dret a l'oblit, perquè el dret pressuposa un objecte —una dada, un enllaç, un responsable— que en el nou règim es dissol.

1 Casanovas (2026), apartat 5.

2 Schacter i Addis (2007).

3 Bühler (1934), § 8. Vegeu Casanovas (2026), apartat 2.

4 Casanovas (2007).

5 Blanchette i Johnson (2002); Mayer-Schönberger (2009), pp. 1–15; Esposito (2017), p. 5.

6 Casanovas (2026), apartat 2.

Procediré en cinc passos. Repassaré què sabem de l'oblit des de la ciència cognitiva, la filosofia i les ciències socials (apartat 2); reconstruiré els orígens, la jurisprudència i les paradoxes del dret a l'oblit (apartat 3); proposaré una periodització de la literatura especialitzada (apartat 4); analitzaré què li passa al dret quan entra en contacte amb la IA (apartat 5); i acabaré preguntant què podem esperar (apartat 6). El fil conductor és el que he desenvolupat en treballs anteriors sobre la governança jurídica i els ecosistemes jurídics i ètics intel·ligents: en una societat híbrida, els drets no es garanteixen només amb normes i sentències, sinó amb el disseny dels entorns on humans i sistemes intel·ligents interactuen.[7]

## 2 El concepte d'oblit en ciència cognitiva, filosofia i ciències socials

### 2.1 Oblit i memòria, individual i col·lectiva

La psicologia experimental va començar mesurant l'oblit abans que la memòria. La corba d'Ebbinghaus (1885) descrivia la pèrdua com una funció del temps; Bartlett (1932) va mostrar que el que queda no és una còpia minvada sinó una reconstrucció guiada per esquemes culturals.[8] Durant un segle, tanmateix, l'oblit fou tractat com una fallada passiva: decaïment o interferència. El canvi de perspectiva és recent i és net. Anderson i Hulbert han documentat mecanismes de control prefrontal que suprimeixen activament records competidors: oblidem, en part, perquè el cervell ho fa activament, i de vegades perquè ho volem.[9] Ja en el Col·loqui de 2022 vaig posar exemples de la impenetrabilitat cognitiva com a protecció, és a dir, de l'oblit com a mecanisme de supervivència.[10] John J. Drummond, per exemple, relata com de petit va caure sobre un niu d'abelles i en va rebre més de setanta picades: la por i el dolor perduren, mentre que els detalls de l'episodi se li han esborrat de la memòria.[11]

La neurobiologia ha mostrat que la funció de la memòria no és transmetre informació fidelment a través del temps sinó optimitzar la decisió en entorns canviants, i que la transitorietat serveix a aquest fi de dues maneres: augmenta la flexibilitat, reduint el pes d'informació caducada, i evita el sobreajustament (*overfitting*) a episodis particulars, afavorint-ne la generalització.[12] El vocabulari és el de l'aprenentatge automàtic, i no per casualitat. Hi tornarem.

La literatura ho sabia. El Funes de Borges, incapaç d'oblidar, era també incapaç de pensar, perquè «pensar es olvidar diferencias, es generalizar, abstraer»; i el mnemonista que Aleksandr Luria va seguir durant trenta anys, Solomon V. Shereshevsky, vivia assetjat per una memòria que no sabia desprendre's de res, sostinguda per imatges que situava mentalment en un ordre concret, com en l'antic mètode dels llocs.[13] Schacter ha sistematitzat

7 Casanovas (2024); Casanovas i Noriega (2026).
8 Ebbinghaus (1885); Bartlett (1932).
9 Anderson i Hulbert (2021).
10 Casanovas (2023).
11 Drummond (2004), pp. 109–110.
12 Richards i Frankland (2017).
13 Borges (1944); Luria (1968).

els «set pecats» de la memòria —transitorietat, distracció, bloqueig, atribució errònia, suggestibilitat, biaix i persistència— com a subproductes de trets adaptatius: el preu d'un sistema que funciona.[14] Retinguem-ne dos, l'atribució errònia i la suggestibilitat, perquè són exactament els que la IA conversacional és capaç d'explotar (apartat 5.3).

La filosofia ha pensat l'oblit com a amenaça i com a condició. Al *Fedre*, el rei Thamus adverteix que l'escriptura és un *phármakon* (remei i verí alhora) que produirà oblit en les ànimes que se'n refiïn; és la primera crítica de l'externalització de la memòria, i l'hem repetida amb cada tecnologia. L'experiència dels aborígens australians, capaços de transmetre memòria i coneixement durant mil·lennis sense escriptura, mitjançant símbols, relats, cants i pictogrames, mostra que una memòria sense escriptura pot ser extraordinàriament duradora, i en aquest sentit sembla donar-li la raó.[15] Nietzsche va invertir-ne el signe: sense la capacitat d'oblidar no hi ha acció ni felicitat, i un excés d'història ofega la vida.[16] Ricœur ha donat la cartografia més completa: distingeix l'oblit per esborrament de traces de l'oblit de reserva —allò que no recordem però resta disponible—, i al costat dels abusos de la memòria (impedida, manipulada, obligada) situa els abusos de l'oblit, dels quals l'amnistia —un oblit prescrit, és a dir, imposat— és la forma institucional.[17] Un precedent clàssic és l'Atenes del 403 aC, on, després de la guerra civil, els ciutadans van jurar *mē mnēsikakeîn*: no recordar els greuges.[18] Renan en farà, el 1882, una condició de la nació: l'oblit, i fins i tot l'error històric, és un factor essencial de la seva creació.[19] Els *lieux de mémoire* de Pierre Nora —les entitats que funcionen com a referents emblemàtics d'una comunitat, d'un grup social o d'una nació— neixen, segons Nora, precisament perquè ja no hi ha *milieux de mémoire*: són la memòria que una societat fixa quan sent que l'està perdent.[20]

Les ciències socials, finalment, han mostrat que ni el record ni l'oblit són mai purament individuals. Fa poc més d'un segle que Halbwachs publicà *Les cadres sociaux de la mémoire* (1925): recordem dins de marcs que proporciona el grup, i quan el marc desapareix el record esdevé inaccessible.[21] Podem afegir-hi, amb Jan Assmann, la distinció entre memòria comunicativa, que lliga tres generacions, i memòria cultural, que requereix suports i

14 Schacter (2001).

15 Plató, *Fedre* 274c–275b. Sobre la transmissió aborigen del coneixement, vegeu Yunkaporta (2019), que en descriu la pràctica del *yarning* —una conversa narrativa, no lineal i orientada a aprendre i a compartir—, i Pascoe (2014).

16 Nietzsche (1874), § 1.

17 Ricœur (2000), pp. 536–589.

18 Loraux (1997). Després de la derrota d'Atenes a la guerra del Peloponès, es va instaurar el règim oligàrquic dels Trenta, recolzat per Esparta, que va desembocar en una guerra civil guanyada pels demòcrates de Trasíbul. Per evitar una espiral de revenges, l'acord de reconciliació del 403 aC va imposar als ciutadans el jurament de μὴ μνησικακεῖν (*mē mnēsikakeîn*), que Andòcides transmet així: «No recordaré els greuges contra cap ciutadà, excepte els Trenta, els Deu i els Onze, i ni tan sols contra aquests si accepten retre comptes del seu càrrec» (Andòcides, *Sobre els misteris* 90). Cf. Carawan (2012), però vegeu la discussió sobre l'autenticitat dels documents a Canevaro i Harris (2012). L'escola d'Aristòtil ho recull de la manera següent: «A ningú no serà permès de retreure el passat a ningú, llevat als Trenta, als Deu, als Onze i als antics magistrats del Pireu, quan hauran retut comptes» (*Constitució d'Atenes* XXXIX.6; Aristòtil 1946, p. 37).

19 Renan (1882).

20 Nora (1984–1992), especialment la introducció, «Entre mémoire et histoire», al vol. I.

21 Halbwachs (1925).

especialistes; i, amb Aleida Assmann, la diferència entre el cànon que una societat manté actiu i l'arxiu que conserva passivament.[22] L'antropòleg social Paul Connerton va desfer la identificació entre oblit (*forgetting*) i fracàs (*failure*) distingint-ne set tipus, de l'esborrament repressiu a l'oblit constitutiu d'una nova identitat.[23] Em pregunto com hauria qualificat el gir que la IA ha imprès en la memòria col·lectiva de les societats modernes, ell que va sostenir que la modernitat té un problema particular amb l'oblit.[24] El diagnòstic es manté, però potser per raons diferents de les que ell adduïa: no tant per la pèrdua de l'escala humana de la vida social com per la delegació de la selecció del que es recorda (apartat 5.3). Potser podríem afegir doncs un vuitè tipus d'oblit a la seva taxonomia, l'oblit algorítmic, sense agent però amb disseny.

## 2.2 Com ha estat tractat el problema: definicions i estat de l'art

La taula 1 ordena aquestes aportacions. Tres idees en resulten, que ens serviran de criteri més endavant.

1. L'oblit és constitutiu, no deficitari; un sistema de memòria que no oblida no és millor sinó disfuncional.
2. Oblidar no equival a destruir. Tulving i Pearlstone ja distingien el 1966 entre la *disponibilitat* d'una informació emmagatzemada i la seva *accessibilitat* en un context de recuperació; la major part del que anomenem oblit és pèrdua d'accés, no de traça.[25] Quan Luciano Floridi, mig segle després, analitzi *Google Spain* amb el mateix parell de conceptes (apartat 3.3), podríem dir que redescobreix per a la infosfera una distinció de la psicologia cognitiva.
3. L'oblit és sempre social, perquè depèn de marcs, suports i decisions col·lectives sobre què mereix romandre en el cànon i què en queda fora.

---

22 J. Assmann (1992); A. Assmann (2016).

23 Connerton (2008). Els set tipus són: (1) l'esborrament repressiu (*repressive erasure*), executat per l'Estat o per un règim autoritari per eliminar de la història oficial fets, imatges o persones incòmodes; (2) l'oblit prescriptiu (*prescriptive forgetting*), un acte de desmemòria decretat o acceptat en interès comú, generalment en transicions polítiques o acords de pau, del qual l'amnistia atenesa és l'exemple clàssic (l'anomenat «pacte de l'oblit» de la Transició espanyola n'és un cas discutit); (3) l'oblit constitutiu d'una nova identitat (*forgetting that is constitutive in the formation of a new identity*), el descart dels records que dificultarien adoptar una narrativa renovada de si mateix; (4) l'amnèsia estructural (*structural amnesia*), inherent a estructures socials com el parentiu, que només retenen allò rellevant per al present, com els avantpassats amb funció genealògica; (5) l'oblit com a anul·lació (*forgetting as annulment*), fruit de la superabundància d'informació i d'arxius, que col·lapsa la capacitat de processar-la; (6) l'oblit com a obsolescència programada (*forgetting as planned obsolescence*), lligat a l'economia de consum, que descarta ràpidament el passat immediat; i (7) l'oblit com a silenci humiliat (*forgetting as humiliated silence*), nascut de la vergonya o d'una ferida col·lectiva, com després d'una derrota o d'una col·laboració.

24 Connerton (2009), pp. 1–6; vegeu també Connerton (1989). Connerton fou un antropòleg proper a les tesis socials de l'Escola de Frankfurt. Així: «What is being forgotten in modernity is profound, the human-scale-ness of life, the experience of living in a world of social relationships that are known. [...] the meaning of life based on shared memories» (2009, p. 5).

25 Tulving i Pearlstone (1966).

**Taula 1** L'oblit en tres tradicions (i en els estudis de memòria digital): nocions i funcions

| **Àmbit** | **Nocions principals** | **Funció atribuïda a l'oblit** | **Referències** |
|---|---|---|---|
| Ciència cognitiva | Decaïment, interferència; oblit actiu (inhibició); disponibilitat *vs.* accessibilitat; memòria constructiva | Adaptativa: flexibilitat, generalització, regulació emocional | Ebbinghaus; Bartlett; Tulving i Pearlstone; Schacter; Anderson i Hulbert; Richards i Frankland |
| Filosofia | *Lēthē*/*alētheia*; *phármakon*; oblit actiu; oblit de reserva *vs.* esborrament; oblit prescrit (amnistia) | Condició de l'acció, del perdó i de la vida en comú; risc d'abús | Plató; Nietzsche; Ricœur; Loraux; Weinrich; Eco |
| Ciències socials | Marcs socials; memòria comunicativa i cultural; cànon i arxiu; llocs de memòria; set tipus d'oblit | Construcció i renovació d'identitats; funció dialèctica; selecció del passat útil | Halbwachs; J. i A. Assmann; Renan; Nora; Connerton |
| Estudis de memòria digital | Ment estesa; cognició distribuïda; memòria connectiva; fi del temps de decaïment; memòria algorítmica | Oblit com a decisió i com a disseny; oblit per multiplicació | Clark i Chalmers; Hutchins; Sparrow et al.; Hoskins; Mayer-Schönberger; Esposito; Stainforth |

Font: elaboració pròpia.

## 2.3 Memòria i entorns digitals

Els processos cognitius i la ment humana no acaben als límits del cervell, sinó que poden abastar eines i objectes de l'entorn. Les tesis de la ment estesa (Andy Clark i David Chalmers) i de la cognició distribuïda (Edwin Hutchins) proporcionen el marc per entendre què passa quan la memòria s'externalitza en dispositius: el quadern de què se serveix Otto per trobar on és el MoMA de Nova York forma part del seu sistema cognitiu, i el pont de comandament d'un vaixell o la cabina de pilotatge d'un avió, tal com els descriu Hutchins, «recorden» coses que cap tripulant no sap per si sol.[26] Més endavant s'ha suggerit experimentalment, amb resultats que les rèpliques posteriors han posat en dubte, que, amb un cercador a mà, recordem menys el contingut i més on trobar-lo: tractaríem internet com una memòria transactiva.[27] En els termes dels meus treballs anteriors, és un cas d'ecologia cognitiva: la unitat d'anàlisi ja no és l'individu sinó l'ecosistema d'agents, artefactes i regles on la cognició s'esdevé.[28]

Els estudis de memòria han descrit aquest pas com un «gir connectiu» (Hoskins): de la memòria col·lectiva, sostinguda per marcs socials, a una memòria connectiva sostinguda per xarxes sociotècniques, on desapareix el «temps de decaïment» (*decay time*) que abans feia la feina de l'oblit.[29] Mayer-Schönberger en va treure conseqüències normatives: si recordar ha esdevingut la norma, cal reintroduir l'oblit, per exemple amb dates de caducitat per a les dades.[30] És el que avui s'anomena oblit per disseny (*forgetting by design*) o, com prefereixo dir, oblit mitjançant el disseny (*forgetting through design*). Tanmateix, com ha argumentat Elena Esposito, no és cert que la xarxa «no oblidi mai»: emmagatzematge i accés són

26 Clark i Chalmers (1998); Hutchins (1995a, 1995b).

27 Sparrow et al. (2011). La rèplica de Camerer et al. (2018) no en va reproduir el resultat principal.

28 Poblet et al. (2019).

29 Hoskins (2011, 2013); Stainforth (2022), p. 258.

30 Mayer-Schönberger (2009), pp. 169–195.

problemes diferents, i la memòria no és magatzem sinó capacitat de seleccionar. Els algorismes no recorden ni obliden, perquè processen dades sense comprendre-les; per això, mentre que el repte de la memòria humana és recordar prou, el de la memòria algorítmica és poder oblidar prou, i de manera controlada.[31] I per això mateix poden eludir la paradoxa que bloquejava tota *ars oblivionalis* —decidir oblidar una cosa obliga a recordar-la— amb l'estratègia que ja apuntava Umberto Eco: no esborrar, sinó multiplicar.[32]

Abans de prosseguir amb les qüestions jurídiques i normatives d'aquesta dimensió tecnològica, he d'introduir breument el que podem anomenar la «dissonància de la política europea». La comissària Viviane Reding proclamava el 2005, en llançar la iniciativa de biblioteques digitals que donaria lloc a Europeana (presentada el novembre de 2008), que sense memòria col·lectiva no som res; el 2010, ja com a vicepresidenta de la Comissió, anunciava la voluntat d'introduir un «dret a l'oblit». La primera iniciativa reposava en una visió estàtica de la memòria com a patrimoni emmagatzemat; la segona arribava com a resposta tardana a unes pràctiques de memòria impulsada per dades que Google ja havia desplegat.[33] Entre el deure de memòria i el dret a l'oblit no hi ha contradicció lògica —tenen objectes diferents—, però sí una tensió institucional que el dret ha hagut d'administrar cas per cas. La sociologia de les traces hi afegeix un matís: el que deixem enrere a la xarxa rarament és un testimoni deliberat; són residus d'interacció que altres recullen i interpreten.[34]

## 3 El dret a l'oblit (*right to be forgotten*)

### 3.1 Orígens i desenvolupament

El dret a l'oblit té tres arrels que convé no confondre.

La primera és *rehabilitadora*: qui ja ha complert condemna té dret a no ser definit perpètuament pel seu passat. És la que inspira la cancel·lació d'antecedents penals, una institució que hem tingut ocasió de formalitzar en el cas australià.[35] És una idea recurrent al segle XX. Es troba a la sentència californiana *Melvin v. Reid* (1931); en el que la doctrina francesa va batejar com a *droit à l'oubli* als anys seixanta; i en el dret a la ressocialització que el Tribunal Constitucional Federal alemany va reconèixer a *Lebach* (1973).[36]

Convé remarcar el paper del cinema i la televisió en aquests orígens. A *Melvin v. Reid*, el film *The Red Kimona* (1925) explotava el passat de Gabrielle Darley, antiga prostituta jutjada i absolta per l'assassinat del seu proxeneta, que havia abandonat aquella vida i s'havia casat el 1919. A França, la pel·lícula de Claude Chabrol *Landru* (1963) va motivar la demanda de Fernande Segret, la darrera amant de l'assassí en sèrie, que encara vivia; la sentència del Tribunal de Grande Instance del Sena (1965), comentada per Gérard Lyon-Caen el 1966, és a l'origen doctrinal de l'expressió *droit à l'oubli*. A Alemanya, la ZDF havia preparat un

31 Esposito (2017), pp. 1–2.

32 Esposito (2017), pp. 6–7; Eco (1988).

33 Stainforth (2022), pp. 257–258, 261.

34 Comunello et al. (2022).

35 Governatori et al. (2020).

36 Singh i Malik (2026), p. 53 (*Melvin v. Reid*); Weber (2011), p. 121 (*Lebach*); sobre els antecedents francesos i italians, Mantelero (2013).

docudrama sobre l'assalt de 1969 a un dipòsit de municions de la Bundeswehr a Lebach, en què van ser assassinats quatre soldats de guàrdia; un dels condemnats, a sis anys de presó per complicitat, estava a punt d'obtenir la llibertat condicional.[37]

La segona arrel és la *protecció de dades*: els principis de limitació de la finalitat i de la conservació, i els drets de supressió i d'oposició presents en la Directiva 95/46/CE (arts. 12 i 14), que no parlen d'oblit sinó de tractament de les dades.

La tercera radica en la protecció de la *dignitat*: els drets de la personalitat —honor, intimitat, pròpia imatge— i, encara abans, el «dret a ser deixat en pau» (*the right to be let alone*), expressió del jutge Thomas M. Cooley que Samuel Warren i Louis Brandeis van popularitzar a «The Right to Privacy» (*Harvard Law Review*, 1890).[38]

Robert Post ha insistit que els dos darrers plantejaments responen a lògiques diferents. La privacitat de les dades és instrumental i burocràtica. La funció dignificadora, en canvi, protegeix normes socials de respecte. La seva tesi és que bona part de la confusió posterior prové d'haver-les fusionat.[39]

La formulació contemporània és europea. La proposta de reglament general de protecció de dades de 2012 va incloure un «dret a l'oblit i a la supressió»; el text final (Reglament (UE) 2016/679, RGPD) el recull a l'article 17 com a «dret de supressió ("dret a l'oblit")», amb sis supòsits habilitants i cinc excepcions, entre les quals la llibertat d'expressió i informació, i els fins d'arxiu en interès públic, de recerca científica o històrica.[40] L'article 17.2 hi afegeix l'obligació, per al responsable que hagi fet públiques les dades, d'informar de la sol·licitud els altres responsables: un primer reconeixement que la informació circula en cadena. A l'Estat espanyol, la Llei orgànica 3/2018 desdobla el dret en dues modalitats —cerques a internet (art. 93) i xarxes socials i serveis equivalents (art. 94)—, amb una protecció reforçada per als menors. Fora d'Europa el panorama és fragmentari: Alnimer hi veu un «món fragmentat» on la pretensió extraterritorial europea xoca amb sobiranies digitals diverses, i proposa, seguint l'enfocament per capes de Svantesson, un sistema multinivell.[41]

### 3.2 La jurisprudència europea i nacional

La taula 2 recull les resolucions principals. Dues línies s'hi entrecreuen. El Tribunal de Justícia de la Unió Europea (TJUE) ha construït, a partir de la protecció de dades, un dret a la desindexació enfront dels cercadors. El Tribunal Europeu de Drets Humans (TEDH), a partir de l'equilibri entre els articles 8 i 10 del Conveni, ha hagut de decidir si l'oblit pot arribar a la font: l'hemeroteca digital.

*Google Spain* (2014) té un origen que ens és proper. El 1998 *La Vanguardia* havia publicat, per mandat legal, dos anuncis de subhasta d'immobles per deutes a la Seguretat

37 *Melvin v. Reid*, 112 Cal. App. 285 (1931); TGI Seine (1965), *JCP* 1966 II 14482, nota de G. Lyon-Caen (el tribunal va desestimar la demanda perquè la mateixa Segret havia publicat les seves memòries); BVerfGE 35, 202 (*Lebach*).

38 Cooley (1888), p. 29; Warren i Brandeis (1890), p. 195.

39 Post (2018), pp. 981–983.

40 Reglament (UE) 2016/679, art. 17 i considerants 65–66.

41 Alnimer (2026), pp. 1–2; vegeu Svantesson (2013).

Social on figurava el nom de Mario Costeja. Digitalitzada l'hemeroteca, una cerca pel seu nom els retornava en primer lloc una dècada després, quan el deute estava saldat. L'Agència Espanyola de Protecció de Dades va desestimar la reclamació contra el diari i va estimar la dirigida contra Google; l'Audiència Nacional va plantejar la qüestió prejudicial. El Tribunal va establir que el gestor d'un motor de cerca tracta dades personals i n'és responsable; que la seva activitat afecta els drets de manera addicional i diferent a la de l'editor, perquè permet obtenir «una visió estructurada» de la persona; i que l'interessat pot exigir la supressió d'enllaços a informació lícita i veraç quan, pel temps transcorregut, resulti inadequada, no pertinent o excessiva, llevat que el paper públic de la persona justifiqui un interès preponderant del públic.[42] La contribució més duradora és procedimental: separar la pervivència de la font de la seva localització pel nom.[43]

**Taula 2** Resolucions principals sobre el dret a l'oblit

| Òrgan i data | Assumpte | Doctrina |
|---|---|---|
| TJUE, 13.5.2014 | C-131/12, *Google Spain* (Costeja) | El cercador és responsable del tractament; dret a la desindexació d'informació lícita però inadequada, no pertinent o excessiva pel pas del temps; límit en el paper públic de l'interessat. |
| TJUE, 9.3.2017 | C-398/15, *Manni* | No hi ha, en principi, dret a l'oblit respecte dels registres mercantils; excepcionalment, accés limitat després d'un termini prou llarg. |
| TJUE, 24.9.2019 | C-136/17, *GC i altres c. CNIL* | Les categories especials de dades i les relatives a condemnes també obliguen el cercador; l'ordenació de resultats ha de reflectir la situació judicial actual. |
| TJUE, 24.9.2019 | C-507/17, *Google c. CNIL* | Abast territorial: desindexació a les versions dels Estats membres, amb mesures que dissuadeixin l'accés des de la UE; el dret de la Unió no imposa, ni prohibeix, l'abast mundial. |
| TJUE, 8.12.2022 | C-460/20, *TU i RE c. Google* | Informació inexacta: desindexació si el sol·licitant aporta proves pertinents i suficients de la inexactitud manifesta, sense necessitat de resolució judicial prèvia; s'estén a les miniatures. |
| TEDH, 16.7.2013 | *Węgrzynowski i Smolczewski c. Polònia* | No correspon als tribunals reescriure la història ordenant retirar un article de l'arxiu digital; és preferible una nota aclaridora. |
| TEDH, 28.6.2018 | *M.L. i W.W. c. Alemanya* | Negar l'anonimització d'arxius de premsa sobre un assassinat notori no vulnera l'art. 8: interès públic persistent i paper dels arxius. |
| TEDH, 25.11.2021 | *Biancardi c. Itàlia* | La responsabilitat d'un editor per no desindexar un article no vulnera l'art. 10; el deure no recau només en els cercadors. |
| TEDH (GS), 4.7.2023 | *Hurbain c. Bèlgica* | Ordenar l'anonimització de l'arxiu digital de *Le Soir* no vulnera l'art. 10; set criteris: naturalesa de la informació, temps, interès actual, notorietat, repercussions, accessibilitat, impacte de la mesura. |
| BVerfG, 5.6.1973; 6.11.2019 | *Lebach*; *Recht auf Vergessen I* i *II* | Dret a la ressocialització; el temps com a factor de ponderació; obligacions graduades dels arxius de premsa; articulació de drets fonamentals nacionals i de la Carta. |
| TS (Sala 1a), 15.10.2015; TC, 4.6.2018 | STS 545/2015; STC 58/2018 (hemeroteca d'*El País*) | L'editor ha d'impedir la indexació per cercadors externs; el TC hi afegeix el cercador intern pel nom; l'arxiu resta inalterat. |
| Cass. (Sez. Un.), 22.7.2019 | Sentència n. 19681 | La reevocació de fets passats és activitat historiogràfica; l'esment del nom requereix un interès públic actual. |

Font: elaboració pròpia a partir de les resolucions i de Kulk i Zuiderveen Borgesius (2018), Askar (2026) i Singh i Malik

42 TJUE (Gran Sala), C-131/12, *Google Spain*, ap. 28–41, 80–81 i 93–99.

43 Singh i Malik (2026), p. 52; Frantziou (2014).

(2026).

La seqüència posterior ha anat delimitant el dret per tres costats. *Per l'objecte*: no abasta en principi els registres públics (*Manni*), sí les dades sensibles i penals (*GC*), i també la informació manifestament inexacta, amb una càrrega probatòria raonable per al sol·licitant (*TU i RE*). *Pel territori*: la desindexació és, per regla general, europea i no mundial (*Google c. CNIL*). I *per la font*: Estrasburg ha passat de la reticència a «reescriure la història» (*Węgrzynowski*) a admetre, a *Hurbain*, que l'anonimització d'una hemeroteca digital pot ser una ingerència proporcionada, tot fixant-ne un test de set criteris.[44] Els tribunals nacionals han explorat solucions intermèdies: a Espanya, la desindexació per part de l'editor sense tocar l'arxiu; a Alemanya, obligacions graduades segons el temps; a Itàlia, la distinció entre crònica i historiografia. El resultat és menys un dret unitari que una *escala de recursos o remeis* (*remedies*) —supressió, desindexació, emmascarament o anonimització, retirada, restricció d'accés— entre els quals cal triar el menys restrictiu que sigui eficaç. Aquesta direcció s'observa també fora d'Europa, per exemple en la sentència del Tribunal Superior de Delhi de 29 de maig de 2026 (*Laksh Vir Singh Yadav c. Union of India*), que distingeix entre preservació, emmascarament i desindexació pel nom en relació amb les resolucions judicials en línia.[45]

### 3.3 Paradoxes, controvèrsies i discussions

Poques figures jurídiques han suscitat un debat tan viu. A favor s'ha argumentat (i) que el dret restaura un control mínim de l'individu sobre la seva identitat informacional i fa efectiu el règim del consentiment (hom ha proposat limitar-lo justament als tractaments consentits); (ii) que protegeix la possibilitat de reinventar-se i manté obert el futur; i (iii) que corregeix una asimetria de poder entre persones i plataformes.[46] En contra, hom hi ha vist (i), abans i tot de la sentència, l'amenaça més gran a la llibertat d'expressió a internet de la dècada que començava, i (ii) un xoc frontal amb la tradició nord-americana de protecció de la informació veraç.[47] Post hi afegeix que *Google Spain* aplica la lògica burocràtica de la privacitat de dades a un actor, el cercador, que sosté l'esfera pública contemporània de la mateixa manera que la premsa sostenia la del segle XIX, i que el Tribunal hauria d'haver raonat des de la privacitat dignificadora, que sí que sap ponderar.[48]

Luciano Floridi, un dels vuit membres independents del consell assessor que Google va

44 TEDH (Gran Sala), *Hurbain c. Bèlgica*, ap. 205; Singh i Malik (2026), pp. 52–53. Els set criteris són: (1) la naturalesa de la informació arxivada; (2) el temps transcorregut des dels fets i des de la publicació inicial i en línia; (3) l'interès actual de la informació; (4) la notorietat de la persona i la seva conducta posterior; (5) les repercussions negatives de la persistència de la informació en línia, com el risc d'uns antecedents penals virtuals perpetus; (6) el grau d'accessibilitat de la informació en l'arxiu digital; i (7) l'impacte de la mesura en la llibertat d'expressió i, en particular, de premsa. Sobre l'origen intern del cas (Cour de cassation belga, 29 d'abril de 2016) i sobre *Biancardi*, vegeu Askar (2026), pp. 457–458.

45 High Court of Delhi, *Laksh Vir Singh Yadav c. Union of India*, W.P.(C) 1021/2016, 29 de maig de 2026, segons Singh i Malik (2026), pp. 50–51, 55.

46 Ausloos (2012); Weber (2011).

47 Rosen (2012), p. 88; Fazlioglu (2013), p. 149.

48 Post (2018), pp. 981–983.

constituir després de la sentència, n'ha ofert la lectura filosòfica. La clau és la distinció entre disponibilitat i accessibilitat: la informació continua publicada, però deixa d'estar enllaçada. En un entorn on accedim als continguts a través d'enllaços, treure l'enllaç equival a amagar el contingut; a la infosfera, diu, el mapa esdevé més important que el territori.[49] És, com hem vist, la distinció de Tulving aplicada a la memòria social. I explica per què la solució és alhora modesta i eficaç: no esborra res, però reintrodueix la constricció que el temps ja no proporciona. El debat ha deixat almenys sis paradoxes ben identificades, a les quals n'afegeixo una setena:

– *La paradoxa de Costeja*. El ciutadà que volia ser oblidat és avui el nom propi d'una sentència que s'estudia a totes les facultats de dret. Floridi observa que, amb els criteris del mateix Tribunal, hom podria sostenir que ha esdevingut un personatge públic i que la informació és ara pertinent.[50]

– *Recordar per oblidar*. Tota decisió d'oblidar assenyala allò que vol ocultar. A partir dels articles desindexats i de dades públiques, un tercer pot redescobrir una part dels enllaços suprimits i identificar-ne alguns sol·licitants.[51] Hi opera una variant de l'anomenat «efecte Streisand», pres de la demanda que l'actriu Barbra Streisand va interposar el 2003 per fer retirar una fotografia de casa seva i que en va multiplicar la difusió: intentar ocultar un assumpte pot despertar la curiositat i multiplicar-ne la repercussió. És una possibilitat real, encara que no generalitzada.

– *La privatització del judici*. La ponderació entre privacitat i interès públic l'efectua en primera instància una empresa privada, amb criteris propis i sense audiència obligatòria de l'editor. L'estudi del mateix Google sobre cinc anys d'aplicació —3,2 milions d'URL sol·licitades; un miler de sol·licitants, el 0,2 % del total, responsables del 16 % de les sol·licituds— il·lustra l'escala i l'opacitat del procés.[52]

– *El nom equívoc*. Com que ningú no pot ser obligat a oblidar, el dret no garanteix l'oblit: genera, en el millor dels casos, una obscuritat pràctica. El nom promet més del que dona i alimenta tant les expectatives dels titulars com les alarmes dels crítics.

– *El territori*. Una desindexació limitada a Europa és eludible; una de mundial exporta a tercers la ponderació europea entre privacitat i expressió.

– *L'individu i l'arxiu*. L'oblit de cadascú afecta la memòria dels altres: la dels implicats en els mateixos fets, la dels historiadors futurs, la de la comunitat. El dret individual s'exerceix sobre un bé que és, en part, comú.[53]

– *La paradoxa de la desidentificació*. Les dades sensibles es publiquen «desidentificades» precisament per protegir-les; però la desidentificació pot fallar, i qui ho demostra pot topar amb la reacció del poder, perquè l'assumpte es polititza de seguida i a ningú no li agrada

49 Floridi (2015), pp. 167–168.

50 Floridi (2015), p. 177.

51 Xue et al. (2016), pp. 395–397, 401.

52 Bertram et al. (2019), p. 959; vegeu també Kulk i Zuiderveen Borgesius (2018).

53 Esposito (2017), p. 3; Stainforth (2022), p. 265.

que el contradiguin o el deixin en evidència. Va passar a Austràlia el 2016–2017. El Departament de Salut va publicar en obert una mostra «desidentificada» de dades de facturació del Medicare Benefits Schedule (MBS) i del Pharmaceutical Benefits Scheme (PBS) corresponent al 10 % de la població. Chris Culnane, Benjamin Rubinstein i Vanessa Teague van mostrar primer que els identificadors xifrats dels proveïdors es podien revertir, i després que els pacients es podien reidentificar sense desxifrar res, encreuant camps no xifrats —dates de procediments mèdics, any de naixement— amb informació externa.[54] Eliminar els identificadors directes (noms, números de Medicare) és insuficient si els quasi-identificadors permeten construir perfils únics. El Govern va retirar el conjunt de dades i va proposar tipificar com a delicte la reidentificació, fet que va desencadenar un debat important sobre la publicació de dades obertes. És un cas que vaig viure de prop: els investigadors, que van trigar molt poc a fer-ho i ho van comunicar al Departament pensant que l'ajudaven, van rebre pressions del mateix Departament de Salut, igual que la Universitat. La desidentificació és la forma institucional de l'oblit de dades, i el cas mostra que aquest oblit tècnic pot fallar i que la reacció institucional no sempre és positiva.

## 4 Estat de l'art en la literatura especialitzada (2009–2026)

Si prenem com a llindar la publicació de *Delete* (2009) i l'anunci de la reforma europea (2010), la literatura es pot ordenar en quatre períodes (taula 3). No són compartiments estancs, però cadascun respon a una pregunta principal, presenta trets distintius i és tractat des de diverses disciplines.

**Taula 3** Periodització de la literatura sobre el dret a l'oblit

| Període | Pregunta principal | Trets | Obres representatives |
|---|---|---|---|
| I. Conceptual i propositiu (2009–2014) | Hauria d'existir? Què és? | Debat normatiu transnacional; genealogia (*oubli*, *oblio*, *Lebach*); propostes tècniques (caducitat) | Mayer-Schönberger (2009); Koops (2011); Weber (2011); Rosen (2012); Ausloos (2012); de Terwangne (2012); Simón Castellano (2012); Mantelero (2013); Fazlioglu (2013) |
| II. Jurisprudencial i de ponderació (2014–2018) | Com es pondera? Qui decideix? | Comentari de *Google Spain*; cercador com a responsable; criteris (temps, paper públic); primers estudis empírics | Frantziou (2014); Pagallo i Durante (2014); Floridi (2015); Sartor (2016); Jones (2016); Xue et al. (2016); Esposito (2017); Kulk i Zuiderveen Borgesius (2018); Post (2018) |
| III. Implementació i abast (2018–2022) | Fins on arriba? Funciona? | RGPD art. 17; territorialitat; hemeroteques; transparència; memòria col·lectiva; primeres alertes sobre aprenentatge automàtic | Fosch-Villaronga et al. (2018); Veale et al. (2018); Bertram et al. (2019); Bourtoule et al. (2021); Stainforth (2022); Comunello et al. (2022) |
| IV. Generatiu (2022–2026) | Pot oblidar una IA? Què vol dir oblidar? | Models de llenguatge, al·lucinacions, *deepfakes*; desaprenentatge; RGPD i Reglament d'IA; nous subjectes (infants, difunts) i geografies; | Hoskins (2024); Cooper et al. (2025); Zhang et al. (2025); Tan (2025); Tutuncuoglu (2025); Reddy et al. (2025); Askar (2026); Fernández Delgado i Cazurro Barahona (2026); Michałkiewicz-Kądziela i |

54 Culnane et al. (2017). La proposta legislativa era la Privacy Amendment (Re-identification Offence) Bill 2016.

| | | |
|---|---|---|
| | memòria algorítmica | Staszewski (2026); Cheriti et al. (2026); Alnimer (2026); Singh i Malik (2026) |

Font: elaboració pròpia. Els períodes se solapen als anys de transició.

El primer període és *conceptual*. La discussió és si el dret ha d'existir, amb un marcat eix transnacional, i la disciplina dominant és el dret constitucional i de la informació.[55] El segon és *jurisprudencial*: la sentència desplaça la pregunta del *si* al *com*, i la filosofia de la informació (Floridi), la teoria del dret (Sartor, amb la seva anàlisi de l'evolució dels interessos en el flux del temps; Pagallo i Durante) i la sociologia (Esposito) s'incorporen al debat.[56] El tercer és *d'implementació*: amb el RGPD aplicable des del 2018, la literatura es torna empírica i comparada, i apareixen els primers treballs que adverteixen que els sistemes d'aprenentatge automàtic «recorden» d'una manera que el dret no havia previst —els humans obliden, les màquines recorden—, així com la demostració que els models poden ser objecte d'atacs d'inversió que en recuperen dades d'entrenament.[57]

El quart període, en què ens trobem ara, és *generatiu*. La irrupció pública dels grans models de llenguatge a finals de 2022 canvia l'objecte. Reddy i col·laboradors documenten bibliomètricament el creixement del desaprenentatge automàtic (*machine unlearning*) com a camp, motivat explícitament pel dret a l'oblit del RGPD.[58] S'hi observen tres desplaçaments:

(i) *Disciplinari*: el centre de gravetat passa del dret constitucional a les ciències de la computació, amb el risc que «oblidar» es redefineixi com allò que és tècnicament mesurable.

(ii) *Subjectiu*: el dret s'estén a titulars que no el poden exercir per si mateixos; per exemple, els infants víctimes de violència sexual digital, per als quals la lentitud del recurs és una forma de revictimització (Irwan et al.), o les persones difuntes, les restes digitals de les quals alimenten ara continguts generats per IA (Al-Mutairi).[59]

(iii) *Geogràfic*: les contribucions de 2025 i 2026 que he pogut consultar venen del Golf, l'Índia, Indonèsia, la Xina, Polònia, Espanya i Algèria. El dret a l'oblit ha deixat de ser una singularitat europea per esdevenir un lloc comú del constitucionalisme digital global, amb totes les divergències que això comporta.

## 5 El dret a l'oblit en contacte amb la intel·ligència artificial

### 5.1 El tercer règim: de la dada localitzable a la representació distribuïda

La figura 1 resumeix l'argument. En el règim de l'*arxiu*, la memòria resideix en documents; l'operació és conservar, i el recurs és suprimir o anonimitzar. En el règim del *cercador*, la memòria rellevant és l'índex; l'operació és recuperar i ordenar, i el recurs és desindexar. En

55 De Terwangne (2012); Simón Castellano (2012). Entre nosaltres, els treballs de Cécile de Terwangne a la revista *IDP* de la UOC i de Pere Simón Castellano en van fixar aviat els termes.

56 Sartor (2016); Pagallo i Durante (2014).

57 Fosch-Villaronga et al. (2018); Veale et al. (2018).

58 Reddy et al. (2025).

59 Irwan et al. (2026), p. 66; Al-Mutairi (2026).

tots dos casos hi ha un objecte discret —un registre, un enllaç— i un responsable identificable. En el règim del *model generatiu*, la informació apresa no s'emmagatzema com a registre sinó que es distribueix en milers de milions de paràmetres; l'operació és inferir i generar; i no hi ha cap «lloc» on anar a buscar la dada per esborrar-la. A la parella disponibilitat/accessibilitat s'hi afegeix un tercer terme, que proposo d'anomenar *generabilitat*: la capacitat d'un sistema de produir informació sobre una persona, sigui o no present en cap font.

| | **I. ARXIU**<br>*Disponibilitat · 1995–2014* | **II. CERCADOR**<br>*Accessibilitat · 2014–2022* | **III. MODEL GENERATIU**<br>*Generabilitat · 2022–* |
|---|---|---|---|
| **On resideix la memòria** | Document, registre, hemeroteca | Índex, rànquing, perfil agregat | Paràmetres, vectors, memòria d'agents |
| **Operació** | Emmagatzemar i conservar | Recuperar i ordenar | Inferir i (re)generar |
| **Solució d'oblit** | Supressió, anonimització, cancel·lació | Desindexació (delisting), obscuritat pràctica | Desaprenentatge, filtres, control de la identitat |
| **Límit de la solució** | Interès històric, llibertat d'informació | Territorialitat, efecte Streisand | No-localitzabilitat, verificació, al·lucinació |

*Del registre discret a la representació distribuïda: l'oblit deixa de ser una operació local*

**Fig. 1** Tres règims de la memòria digital i les seves solucions d'oblit. Font: elaboració pròpia

Que això planteja un problema jurídic, i no només tècnic, es va discutir durant un temps. L'autoritat de protecció de dades d'Hamburg va sostenir el juliol de 2024 que els paràmetres d'un model no són dades personals, perquè no emmagatzemen informació sobre individus concrets. El Comitè Europeu de Protecció de Dades, en el Dictamen 28/2024 (desembre de 2024), ha adoptat una posició més exigent: un model entrenat amb dades personals no pot considerar-se anònim llevat que es demostri que la probabilitat d'extreure'n dades o d'obtenir-ne mitjançant consultes és insignificant.[60] La recerca tècnica li dona suport: es poden extreure de models de llenguatge fragments literals de dades d'entrenament, inclosa informació personal.[61]

El cas que millor il·lustra el desajust és la reclamació que *noyb* (*None of your business* – European Center for Digital Rights), l'associació de Max Schrems, va presentar contra OpenAI davant l'autoritat austríaca de protecció de dades el 2024. El sistema atribuïa a una persona pública una data de naixement incorrecta; requerida a rectificar o suprimir, l'empresa va reconèixer que no podia impedir que el model mostrés la data errònia sense bloquejar tota la informació sobre l'interessat: el filtre només es podia aplicar globalment, no selectivament.[62] Una segona reclamació, el 2025, davant l'autoritat noruega, concernia un ciutadà a qui el sistema atribuïa falsament l'assassinat de dos dels seus fills. Fernández

60 HmbBfDI (2024); CEPD (2024); Fernández Delgado i Cazurro Barahona (2026), p. 3.

61 Carlini et al. (2021).

62 Fernández Delgado i Cazurro Barahona (2026), p. 8; noyb (2024).

Delgado i Cazurro Barahona en conclouen que no som davant d'un buit normatiu sinó d'un problema de compliment, i que les solucions disponibles són, de moment, pal·liatives i no estructurals.[63]

Quina podria ser la solució? El desaprenentatge automàtic és la resposta tècnica. La idea és eliminar la influència d'uns exemples d'entrenament sense reentrenar el model sencer.[64] La recerca n'ha assenyalat els límits amb una claredat que els juristes haurien de tenir present: treure informació del conjunt d'entrenament no garanteix que el model no pugui generar-la, perquè pot inferir-la d'altres dades; suprimir sortides no és esborrar; i els objectius jurídics no es deixen traduir sense residu en mètriques tècniques.[65] Des del dret s'han formulat diverses propostes: interpretar extensivament el concepte de «supressió» (Tan) o, constatant que el Reglament d'IA no preveu cap mecanisme específic d'oblit per als sistemes generatius, crear-ne de supervisió algorítmica, com ara un comitè específic (Askar).[66] Hi afegeixo una observació. El paral·lelisme amb la ciència cognitiva (apartat 2.1) és instructiu però enganyós. Els models ja «obliden» en el sentit de Richards i Frankland —generalitzen, regularitzen, perden els detalls—; el que no fan és oblidar a voluntat d'un tercer i de manera verificable. I el dret a l'oblit no ha consistit mai en l'oblit del sistema, sinó en el poder d'una persona sobre el que el sistema fa amb ella.

### 5.2 De recordar a generar: identitat, al·lucinació i *deepfake*

La segona característica del nou règim és que el dany ja no prové només de la persistència d'informació veraç. *Google Spain* tractava de fets certs que el temps havia tornat impertinents. Els sistemes generatius produeixen enunciats sobre persones que mai no han estat certs —al·lucinacions— i imatges, veus i vídeos de fets que mai no han ocorregut. Davant dels *deepfakes*, l'article 17 perd eficàcia: pressuposa una dada preexistent que es pugui identificar i suprimir, mentre que el contingut sintètic es crea i es replica més de pressa que qualsevol procediment. Michałkiewicz-Kądziela i Staszewski proposen, per això, redefinir el dret com un dret a controlar la pròpia identitat digital, amb mecanismes preventius *ex ante* i no només reactius.[67]

La proposta és discutible en els detalls, però encerta el diagnòstic: l'objecte del dret s'ha desplaçat de la informació *sobre* mi a la capacitat de *representar-me*. Això connecta el dret a l'oblit amb l'arrel dignificadora que Post reivindicava, i amb dos supòsits límit que la literatura recent ha identificat. El dels difunts, les traces dels quals alimenten continguts generats per IA —i, cada cop més, avatars conversacionals—: qui exerceix l'oblit de qui ja no hi és, i enfront de qui?[68] I el de les neurodades, on el que s'infereix no és el que hem fet sinó

63 Fernández Delgado i Cazurro Barahona (2026), pp. 8–9. Sobre la segona reclamació, presentada davant l'autoritat noruega (Datatilsynet), vegeu noyb (2025).

64 Cao i Yang (2015); Bourtoule et al. (2021).

65 Cooper et al. (2025).

66 Tan (2025), p. 60; Askar (2026), pp. 451, 469–472.

67 Michałkiewicz-Kądziela i Staszewski (2026), pp. 161–163.

68 Al-Mutairi (2026).

el que pensem o sentim.[69]

### 5.3 La prefiguració de la memòria col·lectiva

Fins aquí, el problema individual. Però la IA no només dificulta que algú sigui oblidat; reconfigura el que una societat recorda. Es coneix com a *memòria algorítmica* el procés pel qual els sistemes d'IA, en decidir què s'emmagatzema, què emergeix i què s'esborra, no es limiten a processar informació sinó que construeixen memòria en tres nivells: personal, institucional i social.[70] Cheriti i col·laboradors qualifiquen la IA de «fabricant silenciós» de la memòria col·lectiva i d'«editor invisible» de la història dels mitjans: algorismes que suprimeixen material políticament sensible, cercadors que enfosqueixen continguts comercialment inconvenients, eines automàtiques que privilegien uns materials d'arxiu sobre uns altres.[71] L'oblit, aquí, no és un dret que s'exerceix sinó un efecte que es pateix. Connerton hi hauria reconegut trets de l'esborrament repressiu, però sense cap censor identificable; s'acosta més al que ell anomenava amnèsia estructural, o a l'oblit per anul·lació fruit de la superabundància.

Tres mecanismes em semblen especialment rellevants.

1. El primer és la *selecció*: la jerarquia de resultats, recomanacions i resums fa de cànon, en el sentit d'Aleida Assmann, i relega la resta a un arxiu que ningú no consulta.
2. El segon és la *regeneració*. Spennemann ha demanat a un model de llenguatge (ChatGPT-4o) que imagini el patrimoni cultural de la humanitat d'aquí a trenta anys, en un exercici que és un equivalent contemporani de les postals de Côté que comentàvem l'any passat; i la revisió de Sui i col·laboradors mostra que la restauració generativa del patrimoni —completar el fresc, reconstruir el monument— és ja una pràctica corrent que planteja problemes de coherència cultural i autenticitat.[72] El passat que veurem serà, en part, un passat sintetitzat.
3. El tercer és la *retroalimentació*. Experiments recents indiquen que un interlocutor basat en models de llenguatge indueix significativament més falsos records que un qüestionari, i que imatges i vídeos editats amb IA poden implantar records d'esdeveniments no viscuts: la suggestibilitat i l'atribució errònia de Schacter, industrialitzades.[73] I a escala col·lectiva, Shumailov i col·laboradors han demostrat que els models entrenats recursivament amb dades generades per altres models col·lapsen: obliden primer les cues de la distribució (*tails*), és a dir, allò infreqüent, minoritari, singular.[74] Per a una llengua i una cultura de dimensió mitjana com la nostra, això no és una curiositat tècnica. És una forma d'oblit estructural que no demana cap decisió de ningú.

---

69 Fernández Delgado i Cazurro Barahona (2026), pp. 4, 7.

70 Tutuncuoglu (2025), p. 1.

71 Cheriti et al. (2026), pp. 26, 32.

72 Spennemann (2026); Sui et al. (2026), pp. 437, 443–444.

73 Chan et al. (2024); Pataranutaporn et al. (2025).

74 Shumailov et al. (2024).

Aquí és on el fil amb la lliçó de l'any passat es fa visible. Pedreschi i col·laboradors descriuen la coevolució entre humans i IA com un bucle on les decisions dels usuaris generen dades que entrenen models que configuren les preferències posteriors dels usuaris.[75] Substituïu «preferències» per «records» i tindreu la prefiguració de la memòria. Entre els riscos de la societat híbrida hi havia aleshores quatre formes de relaxació: pràctica, epistèmica, ètica i simbòlica. Caldria afegir-n'hi una cinquena, la *relaxació mnemònica*: la delegació a les màquines no ja del record —això ho fem des de l'escriptura— sinó de la selecció del que val la pena de recordar i d'oblidar. La *Deixis am Phantasma* digital de què parlava, la possibilitat d'entrar dins la nostra imaginació i estatjar-nos-hi, val també per al passat: podem acabar habitant una memòria que no hem format nosaltres. Això es documenta per a les identitats diaspòriques, on les narratives digitals poden tant preservar la memòria col·lectiva com soscavar-la.[76]

### 5.4 De la solució individual a la governança de l'ecosistema

Si l'anàlisi és correcta, el dret a l'oblit no es pot garantir en el tercer règim amb l'instrumental dels dos primers: una sol·licitud, un responsable, una ponderació, un acte d'esborrament o desindexació. No perquè aquest instrumental sigui erroni —continua essent necessari per a arxius i cercadors, i per a la capa de recuperació dels sistemes d'IA— sinó perquè és insuficient. En treballs anteriors he definit la *governança jurídica* com el conjunt de processos que generen un ecosistema regulador sostenible que reflecteixi els conceptes fonamentals d'una democràcia; els *ecosistemes jurídics* com a sistemes complexos i dinàmics amb múltiples nivells i actors; i els *ecosistemes jurídics intel·ligents* com a sistemes reguladors parcialment incrustats en sistemes ciberfísics, orientats al compliment en temps real.[77] El model distingeix quatre conglomerats normatius —dret imperatiu, polítiques, dret tou i ètica — i situa la validesa al centre.[78] L'enfocament no és ni de dalt a baix ni de baix a dalt, sinó *des del mig* i *des de dins* (*middle-out, inside-out*): les regles s'articulen en el disseny mateix dels sistemes.

75 Pedreschi et al. (2025).

76 Kianpour et al. (2026), pp. 36, 42.

77 Casanovas i Noriega (2026); Casanovas et al. (2022); Casanovas et al. (2024).

78 Casanovas (2025d); Casanovas, Pastor Sempere i Echebarría Sáenz (2026).

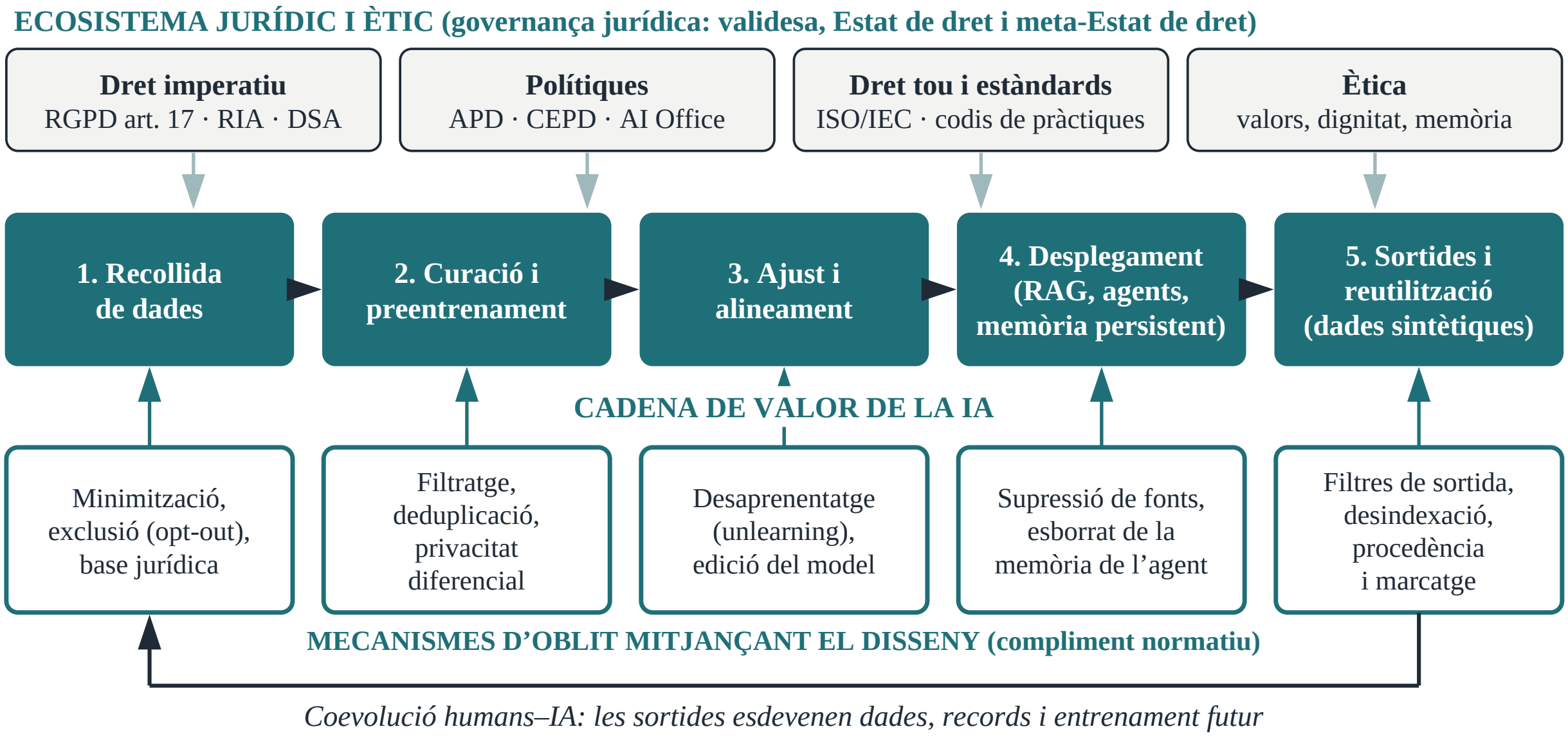

**Fig. 2** El dret a l'oblit distribuït al llarg de la cadena de valor de la IA dins d'un ecosistema jurídic i ètic. Font: elaboració pròpia a partir de Casanovas (2024, 2025f) i Casanovas i Noriega (2026)

Aplicat al nostre problema (fig. 2), això vol dir tres coses. Primera: l'oblit s'ha de *distribuir* al llarg de la cadena de valor de la IA. A cada baula hi correspon un mecanisme diferent —minimització i exclusió en la recollida; filtratge, deduplicació i privacitat diferencial en l'entrenament; desaprenentatge i edició en l'ajust; supressió de fonts i esborrat de memòries persistents en el desplegament; filtres, desindexació i procedència en les sortides— i un responsable diferent. He argumentat en un altre lloc que la cadena de valor, concepte econòmic que el legislador europeu ha convertit en jurídic, ha d'incorporar les dimensions intangibles —llengua, cultura, valors ètics i jurídics— que no són directament monetitzables;[79] la memòria n'és una. Segona: l'oblit s'ha de garantir *mitjançant el disseny*, com a cas particular del compliment normatiu mitjançant el disseny, i s'ha de poder *validar* empíricament, no només declarar. Tercera: cal preservar el que aquest dret té d'específicament jurídic. Una ponderació entre la dignitat d'una persona i la memòria d'una comunitat no és automatitzable; el que sí que es pot dissenyar és l'entorn on aquesta ponderació es fa amb garanties, amb traçabilitat i amb un humà que en respon.[80] Convé dir-ho ara que el debat europeu sobre simplificació reguladora podria afeblir les bases jurídiques del tractament de dades per a l'entrenament de models: la falsa elecció entre regulació i innovació es paga, entre altres monedes, en capacitat d'oblit.[81]

## 6 Conclusions: què podem esperar?

Resumeixo l'argument en sis tesis i tres expectatives.

1. *L'oblit és una funció, no una interrupció del sistema.* La ciència cognitiva, la filosofia i les ciències socials convergeixen a considerar-lo constitutiu de la memòria

79 Casanovas (2025f); vegeu també Casanovas (2025a).

80 Casanovas et al. (2023).

81 Casanovas (2025e).

individual i col·lectiva. Un entorn que no oblida no és més fidel; és disfuncional.

2. *El dret a l'oblit és una escala de recursos, no un dret unitari.* Té tres arrels —rehabilitadora, de protecció de dades i dignificadora— i la jurisprudència l'ha concretat en supressió, desindexació, anonimització i restricció d'accés, a triar segons proporcionalitat. El nom promet més del que dona; el que dona —constricció, obscuritat pràctica— és valuós.
3. *Fou concebut per a dos règims de memòria, l'arxiu i el cercador, i ara opera en un tercer.* A l'arxiu i al cercador hi ha un objecte localitzable i un responsable. Al model generatiu la informació és distribuïda, inferida i regenerable. A la disponibilitat i l'accessibilitat s'hi afegeix la generabilitat.
4. *L'objecte del dret es desplaça de la informació sobre la persona a la capacitat de representar-la.* Al·lucinacions, *deepfakes*, avatars pòstums i neurodades apunten a un dret al control de la identitat digital, que recupera l'arrel dignificadora.
5. *La IA prefigura la memòria col·lectiva per selecció, regeneració i retroalimentació.* L'oblit estructural de les cues de la distribució —allò minoritari i singular— és un risc específic per a llengües i cultures com la nostra. El dret individual a l'oblit i el deure col·lectiu de memòria no són contradictoris: tots dos s'oposen a una memòria administrada sense responsable.
6. *L'exercici efectiu del dret requereix passar de la solució individual* ex post *a la governança de l'ecosistema*: oblit distribuït al llarg de la cadena de valor, garantit mitjançant el disseny, validable empíricament, i amb una instància humana i pública de ponderació.

Què podem esperar? A curt termini, que els tribunals europeus es pronunciïn sobre l'aplicació de l'article 17 als models generatius, probablement estenent la lògica de *Google Spain*: qui determina fins i mitjans respon, i la dificultat tècnica no eximeix. A mitjà termini, una convergència imperfecta entre el dret i la tècnica: el desaprenentatge no satisfarà mai del tot l'ideal de l'esborrament, de la mateixa manera que la desindexació no satisfeia el de l'oblit, però pot arribar a proporcionar una obscuritat suficient si va acompanyat de mecanismes de verificació. A llarg termini, la pregunta és política: qui decideix què recorda i què oblida una societat híbrida? La pitjor resposta és que no ho decideixi ningú, perquè aleshores els poders fàctics, tant polítics com corporatius, poden ocupar aquest lloc i decidir per nosaltres, sense haver de donar-ne comptes.[82]

Acabava l'any passat amb la imatge de les onades gegants, que cap mariner que les hagi vistes no ha sobreviscut per explicar: ens les hem d'imaginar. L'oblit s'hi assembla. Per definició no el podem veure; només en percebem les vores, els límits borrosos, quan busquem alguna cosa i no hi és, o quan trobem el que hauria d'haver desaparegut. Junceda feia exclamar als seus personatges de 1875, 1925 i 1975: «Ah! Oh! Un auto!», «Ah! Oh! Un cavall!», «Ah! Oh! Un que va a peu!». No és difícil imaginar la vinyeta següent. Any 2075: «Ah! Oh! Un que oblida!». Que això resulti una raresa o un dret dependrà del que dissenyem

82 Sobre aquests riscos, vegeu Casanovas (2025b, 2025c).

ara.

## Agraïments

Aquest treball s’emmarca parcialment en les activitats de la unitat associada de la UAB a l’IIIA-CSIC, Research Group on AI and Law (2023–2026), i en el projecte del CSIC CAI – Civic Artificial Intelligence: Artificial Intelligence at the Service of Society (ILINK25174). En la preparació d’aquest text he fet servir Claude (Anthropic) com a assistent per a la bibliografia, la verificació de fonts, l’edició i la confecció de taules i diagrames, i n’he aprofitat també alguns suggeriments. La responsabilitat del contingut és exclusivament meva.

## Referències

## Jurisprudència i fonts normatives